\documentclass[10pt,journal,twoside,final]{IEEEtran}

\usepackage{graphicx}
\graphicspath{{./Pictures/}} 
\usepackage{amsfonts,amssymb} 
\usepackage{dsfont}
\usepackage{tablefootnote}
\usepackage{amsmath}
\usepackage{url}
\usepackage{mathrsfs}
\usepackage{amssymb}

\usepackage{multirow}
\usepackage{array}
\newcolumntype{L}[1]{>{\raggedright\let\newline\\\arraybackslash\hspace{0pt}}m{#1}}
\newcolumntype{C}[1]{>{\centering\let\newline\\\arraybackslash\hspace{0pt}}m{#1}}
\newcolumntype{R}[1]{>{\raggedleft\let\newline\\\arraybackslash\hspace{0pt}}m{#1}}

\usepackage{bigstrut}
\usepackage{cite}

\ifCLASSOPTIONcompsoc
\usepackage[caption=false,font=normalsize,labelfont=sf,textfont=sf,subrefformat=parens,labelformat=parens]{subfig}
\else
\usepackage[caption=false,font=footnotesize,subrefformat=parens,labelformat=parens]{subfig}
\fi

\usepackage{color,xcolor,colortbl}
\newcommand{\etal}{\textit{et al. }}
\newcommand{\mj}[1]{{\color{black}#1}}

\usepackage{algorithm}
\usepackage{algpseudocode}
\algnewcommand\algorithmicinput{\textbf{INPUT:}}
\algnewcommand\INPUT{\item[\algorithmicinput]}
\algnewcommand{\algorithmicoutput}{\textbf{OUTPUT:}}
\algnewcommand\OUTPUT{\item[\algorithmicoutput]}
\algrenewcommand{\algorithmiccomment}[1]{\hskip3em$\%$ #1}

\begin{document}

\title{Towards Channel-Robust and Receiver-Independent Radio Frequency Fingerprint Identification}

\author{Jie~Ma,~\IEEEmembership{Student Member,~IEEE,}
	Junqing~Zhang,~\IEEEmembership{Senior~Member,~IEEE,}
        Guanxiong~Shen,~\IEEEmembership{Member,~IEEE,}
        Linning~Peng,~\IEEEmembership{Member,~IEEE,}
	Alan~Marshall,~\IEEEmembership{Senior Member,~IEEE}
	
\thanks{Manuscript received xxx; revised xxx; accepted xxx. Date of publication xxx; date of current version xxx. The work of J. Zhang was supported in part by the UK Engineering and Physical Sciences Research Council (EPSRC) under grant ID EP/Y037197/1 and in part by the UK Royal Society Research Grants RGS$\backslash$R1$\backslash$231435. This article was presented in part at the IEEE WCNC 2025~\cite{jie2025wcnc}.
For the purpose of open access, the authors have applied a Creative Commons Attribution (CC BY) licence to any Accepted Manuscript version arising.
    The review of this paper was coordinated by xxx.  
\textit{(Corresponding author: Junqing Zhang.)}}
\thanks{J. Ma, J.~Zhang, and A.~Marshall are with the School of Computer Science and Informatics, University of Liverpool, Liverpool, L69 3DR, United Kingdom. (email: Jie.Ma@liverpool.ac.uk; Junqing.Zhang@liverpool.ac.uk; Alan.Marshall@liverpool.ac.uk)}
\thanks{G.~Shen and L.~Peng are with the School of Cyber Science and Engineering, Southeast University, Nanjing, China. (email: gxshen@seu.edu.cn; pengln@seu.edu.cn)}
\thanks{Color versions of one or more of the figures in this paper are available online at http://ieeexplore.ieee.org.}
\thanks{Digital Object Identifier xxx}
}

\maketitle	

\begin{abstract}
Radio frequency fingerprint identification (RFFI) is an emerging method for authenticating Internet of Things (IoT) devices. RFFI exploits the intrinsic and unique hardware imperfections for classifying IoT devices. Deep learning-based RFFI has shown excellent performance. However, there are still remaining research challenges, such as limited public training datasets as well as impacts of channel and receive effects. 
\mj{In this paper, we proposed a three-stage RFFI approach involving contrastive learning-enhanced pretraining, Siamese network-based classification network training, and inference.}
Specifically, we employed spectrogram as signal representation to decouple the transmitter impairments from channel effects and receiver impairments. We proposed an unsupervised contrastive learning method to pretrain a channel-robust RFF extractor. In addition, the Siamese network-based scheme is enhanced by data augmentation and contrastive loss, which is capable of jointly mitigating the effects of channel and receiver impairments. We carried out a comprehensive experimental evaluation using three public LoRa datasets and one self-collected LoRa dataset. The results demonstrated that our approach can effectively and simultaneously mitigate the effects of channel and receiver impairments. We also showed that pretraining can significantly reduce the required amount of the fine-tuning data. Our proposed approach achieved an accuracy of over 90\% in dynamic non-line-of-sight (NLOS) scenarios when there are only 20 packets per device.
\end{abstract}

%


\section{Introduction}

\IEEEPARstart{T}{he} increasing use of wireless devices has raised security concerns regarding device authentication in the Internet of Things (IoT)~\cite{zhang2025tifs}.
Conventional authentication methods rely on cryptographic algorithms combined with software addresses. 
However, cryptographic approaches rely on complicated mathematical algorithms, which tend to be computationally expensive. Many IoT devices are only equipped with limited processing capability and cannot afford costly cryptographic schemes~\cite{zhang2020new}. Moreover, software addresses, e.g., media access control (MAC) addresses, can be easily tampered with. Therefore, there is a strong need for a low-complexity and reliable authentication scheme that is applicable to low cost IoT devices. 

Radio frequency fingerprint identification (RFFI) is a lightweight physical layer-based device authentication technique~\cite{zhang2025tifs}. 
It identifies wireless devices by analyzing the unique distortion of the received waveform caused by the imperfections of hardware components. The imperfection comes from the transmitter chain of the wireless device, such as mixers, oscillators, power amplifiers, etc~\cite{zhang2021radio}. These components deviate slightly from their nominal specifications due to the inevitable manufacturing process variations.
It manifests as hardware imperfections such as I/Q imbalance, carrier frequency offset (CFO), and power amplifier non-linearity~\cite{yao2024novel, jiang2023rf, elmaghbub2024distinguishable, li2022radio,hanna2020open, sankhe2019no, he2024diff, yang2023led}. 

An RFFI system typically involves a few transmitters, also known as devices under test (DUTs), and one or multiple receivers. A receiver will capture the physical layer waveform with the hardware impairments embedded. It will then predict the device identity based on the wireless signals by extracting unique features, using an RFFI engine.
\mj{Deep learning has been the core of the RFFI engine to extract the discriminative features from interrelated hardware impairments\cite{rfdna, shen2023length, xie2021generalizable, qian2021specific, rajendran2022rf, wang2022radio, he2023channel, he2024radio, tang2024causal, cistccn_2024, yin2024multi, zhang2024multisource, peng2024hybrid, li2023lte}. 
In the last decade, deep learning has demonstrated excellent capabilities in feature extraction and classification, leading to significant progress in image processing, natural language processing, etc~\cite{chen2020simple}.
However, RFFI tackles RF signals, which are different from images, videos, or text~\cite{zhou2025tifs}.
There are unique challenges to deep learning-based RFFI, such as limited public datasets~\cite{shen2024federated}, varying channel effects~\cite{linning2024, wang2025twc}, and the hardware impairments of the receiver~\cite{li2025jsac, zhao2024}, etc.}

Deep learning training is data-hungry, which requires a large amount of labelled training datasets.
Acquiring a sufficient amount of training signals requires substantial collection and labour costs.
It is common to fine-tune a pretrained model with a similar task in the deep learning community. However, most of the RFFI literature trains deep learning models from scratch~\cite{rfdna, shen2023length, xie2021generalizable, qian2021specific, rajendran2022rf, wang2022radio, he2023channel, he2024radio, tang2024causal, cistccn_2024, yin2024multi, zhang2024multisource, peng2024hybrid, li2023lte}, with a few exceptions~\cite{xu2024enhanced, shen2022towards, shen2024federated}.
Xu~\etal ~\cite{xu2024enhanced} constructs a pretrained encoder using an unsupervised learning method, but overlooked several key factors, such as channels and receivers.
Shen~\etal ~\cite{shen2022towards} pretrained an RFF extractor in a supervised learning manner, but such a labelled training dataset may not always be available. The work in~\cite{shen2024federated} adopted federated learning to train an RFF extractor among several edge receivers.
However, distributed training may not be necessary when centralized training can be done offline, which will be less complicated.

The varying channel effects will impact the RFF extraction because the multipath will distort the received signals~\cite{al2020exposing}. 
It is very common that RFFI will be subject to various channel effects, especially in mobile IoT systems, which calls for channel mitigation approaches.
Such approaches can be categorized into signal processing-based methods and deep learning-based methods. Signal processing-based methods carefully design algorithms to eliminate channel effects, but preserve hardware impairments. For example, the work in \cite{shen2022towards} proposed channel independent spectrogram, which mitigates the channel effect by dividing the adjacent columns of the spectrogram. Channel equalization is also adopted for RFFI~\cite{al2020exposing}. However, these algorithms may accidentally erase some hardware impairments. In addition, designing such algorithms require expert understanding on the signal features, which is not trivial.
Regarding deep learning-based methods, data augmentation is widely employed~\cite{shen2022towards,soltani2020more}. Specifically, data augmentation enriches the channel effects by simulation in the training stage, to make the training dataset as generalizable as possible. While the effectiveness of data augmentation has been demonstrated, it is not possible to emulate all the possible channel effects that will occur in the test stage.

The receiver hardware impairments also affect the RFFI performance, when multiple receivers are involved and/or different receivers are used in the deep learning training and test~\cite{shen2023towards, zhao2024, zha2023cross,zhang2024domain, li2025jsac}.
The work in~\cite{shen2023towards} incorporates adversarial training to learn receiver-independent features by utilizing the gradient reversal layer to mitigate the influence of receivers. 
Zhao~\etal ~\cite{zhao2024} employs generative adversarial networks (GANs) to address the receiver-related challenges. 
However, a substantial volume of labelled datasets is essential for developing the system, particularly in the first phase, involving over one million signals, along with $50$ transmitters and $28$ receivers. 
Zha~\etal ~\cite{zha2023cross} proposes a cross-receiver RFF learning based on contrastive learning. 
However, the evaluation is limited as only two types of USRP software-defined radio (SDR) platforms were involved as receivers.
\mj{In~\cite{li2025jsac}, a global domain adaptation approach based on adversarial training, along with a subdomain adaptation method leveraging Local Maximum Mean Discrepancy (LMMD), is introduced to facilitate the extraction of receiver-independent features. Zhang \textit{et al.}~\cite{zhang2024domain} proposes a separability condition aimed at minimizing the classification error probability at a new receiver. }

Although previous studies have addressed channel and receiver challenges in RFFI, their effects are mitigated separately. In practice, these effects are highly likely to exist in the same system, e.g., in a mobile IoT system with multiple receivers. Joint mitigation of channel and receiver effects is essential but currently missing.

\mj{This paper proposed a three-stage approach involving contrastive learning-based pretraining, Siamese network-based RFFI classification network training, and inference, in order to tackle the challenges posed by limited training datasets as well as channel and receiver effects. }
We first pretrained an RFF feature extractor using unsupervised contrastive learning. Then, we fine-tuned a Siamese network-based RFFI classification network to jointly mitigate channel and receiver variations.
\mj{During the inference stage, the received samples are fed into the trained neural network for device identification.}
Extensive experimental evaluations have been carried out by using three public LoRa datasets~\cite{shen2022towards,2021access,shen2023towards}. Additionally, we constructed a testbed consisting of ten LoRa DUTs and six SDR platforms as receivers; we collected a dataset that includes dynamic channel variations and multiple receivers. 
We carried out comprehensive experimental evaluations to assess the performance of pretraining and classification.
Our technical contributions are summarized as follows.
\begin{itemize}
\item We used the spectrogram as the signal representation, which decouples transmitter impairments, channel effects, and receiver impairments without resulting in information loss.
 \item We pretrained an RFF feature extractor using unsupervised contrastive learning and publicly available training datasets. Since it is unsupervised learning, the label space is not required, which allows us to freely choose public resources. Data augmentation is adopted to train a channel-robust feature extractor.
 \item  We designed a Siamese network-based RFFI classification network, enhanced by data augmentation and contrastive loss. We specifically constructed two fine-tuning datasets from two receivers for all the DUTs. A pair of these two signals from the same DUT but different receivers were augmented by channel effects. The contrastive learning then jointly mitigates the channel and receiver effects.
 \item We carried out extensive experimental evaluations on three labelled datasets for training/fine-tuning and one unlabelled dataset for pretraining. The experiment results demonstrated that pretraining is effective, especially when the fine-tuning data is limited. In addition, the channel and receiver effects can also be eliminated effectively by the Siamese network-based RFFI. 
\end{itemize}
In our previous work~\cite{jie2025wcnc}, we showed that the contrastive learning-based RFFI system is robust to channel variation. Moreover, compared with IQ signals, spectrogram, and channel-independent spectrogram, the spectrogram with contrastive learning showed a higher accuracy in both static and dynamic channels. 
This paper further extended our previous work by jointly solving the impacts of channel and receivers via a Siamese network with contrastive learning.
In addition, we explored the importance of the pretrain model in scenarios with a limited number of training samples.
We also carried out a more comprehensive experimental evaluation.

The rest of the paper is organized as follows. Section~\ref{sec:related_work} introduces the background and problem statement.
Section~\ref{sec:system} outlines our proposed system. Then, the detailed design methods of the Siamese-based RFFI and contrastive learning-based pretraining are elaborated in Sections~\ref{sec:siameseRFFI} and~\ref{sec:contrasRF}, respectively.
The experimental setup and involved datasets are introduced in Section~\ref{sec:setup}. 
Sections~\ref{sec:eva_siamese} and \ref{sec:eva_pre} present the experimental results for the classification and pretraining, respectively.
Finally, Section~\ref{sec:conclusion} concludes the paper.

\section{Background and Problem Statement}\label{sec:related_work}
\subsection{LoRa Primer}
LoRa is a widely used wireless technique for long-range IoT applications.
LoRa employs chirp spread spectrum (CSS) modulation and the signal's instantaneous frequency varies continuously over time.  
There are several preambles at the beginning of each LoRa packet. 
A baseband LoRa preamble can be mathematically given as
\begin{equation}\label{equ:chirp_rfband}
s(t)=A e^{j\left(-\pi B t+\pi \frac{B}{T} t^2\right)} \quad(0 \leq t \leq T),
\end{equation}
where $A$, $B$, $T$ are the amplitude, bandwidth, and symbol duration, respectively.

\begin{figure}[!t]
	\centering
	\subfloat[]{\includegraphics[width=1.7in]{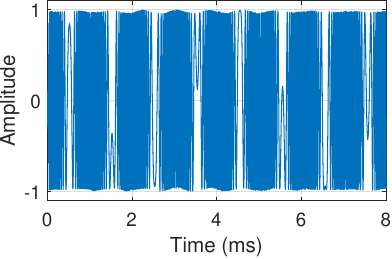}    
		\label{fig:chirpall}}
    \subfloat[]{\includegraphics[width=1.7in]{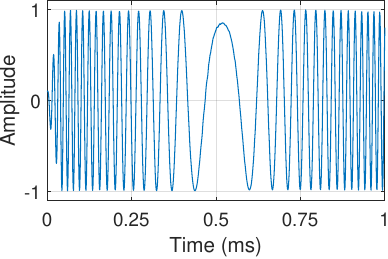}    
		\label{fig:chirp1}}

    \subfloat[]{\includegraphics[width=1.7in]{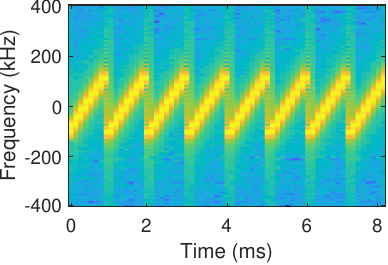}   
		\label{fig:spec}}
    \subfloat[]{\includegraphics[width=1.7in]{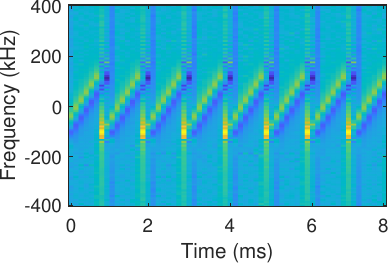}
 \label{fig:cis}}	
 \caption{(a) Time domain signal of the preambles (I branch).  (b) Time domain signal of the first preamble (I branch). (c) Spectrogram of the preambles. (d) Channel-independent spectrogram of the preambles.}
	\label{fig:waveform}
\end{figure}

The contents of the preambles are identical and fixed across all packets, making them ideal for RFFI. Fig.~\ref*{fig:waveform}(a) shows the time-domain baseband signal of the preambles (I branch) and Fig.~\ref*{fig:waveform}(b) zooms the first preamble. 

\subsection{Deep Learning-Based RFFI}
\begin{figure}[!t]
	\begin{center}
		\includegraphics[width = 3.4in]{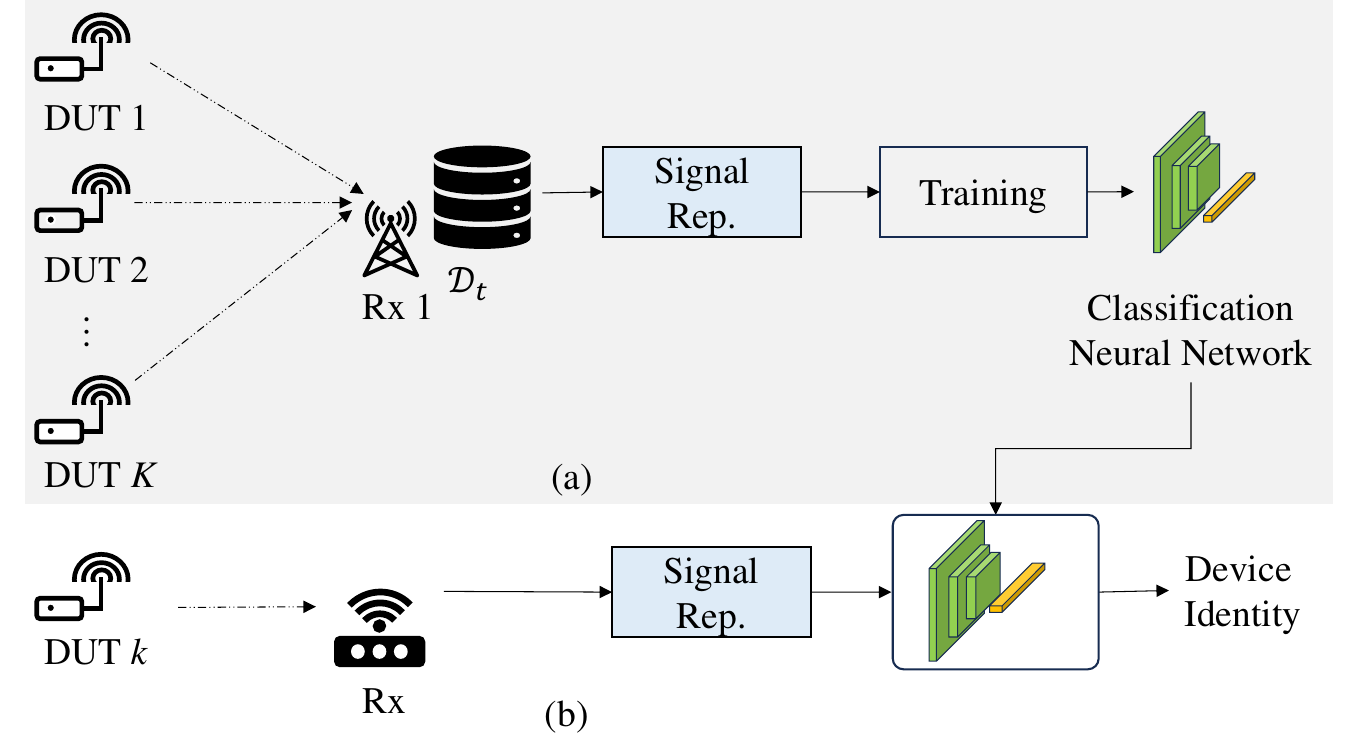}
		\caption{Deep learning-based RFFI. (a) Training stage. (b) Inference stage.}
		\label{fig:convRFF}
	\end{center}	
\end{figure}

As shown in Fig.~\ref{fig:convRFF}, a common deep learning-based RFFI system comprises $K$ DUTs and a receiver. Based on the received signal, the receiver aims to predict the identity of the DUT.
The deep learning-based approach consists of the training and inference stages.
\begin{itemize}
    \item Training stage: A training dataset, i.e., 
    $\mathcal{D}_t=\left\{\left(x_i, y_i\right)\right\}, \; i=1, \ldots, KM,$ 
    is constructed, where $x_i$ and $y_i$ represent the time domain signals (IQ samples) and its label (the index of the DUT), respectively, $M$ is the number of samples from each DUT. The signal is first converted to a particular representation and then fed into a deep learning model for training. A trained model will be returned once the process is completed.
    \item Inference stage: The received signal undergoes the same signal representation procedure as that in the training stage. It is then input into the trained model for classification, and a prediction will be made.
\end{itemize}

The above process is general to deep learning-based RFFI. However, there are a few challenges in practical applications. This paper will focus on the channel and receiver effects as well as the requirement of labelled training data, which will be elaborated below. 

\subsection{Channel and Receiver Effects}\label{sec:channel_rx}
The received signal is expressed as
\begin{equation}
\begin{aligned}
x(t) & = \mathcal{G} * h(t) * \mathcal{F}_k(s(t))+n(t),
\label{eq:time_signal}
\end{aligned}
\end{equation}
where $\mathcal{G}$ represents the effects of the hardware impairments of the receiver, $h(t)$ is the wireless channel impulse response, $*$ represents the convolutional operation, $\mathcal{F}_k(\cdot)$ denotes the effect of the $k^{th}$ DUT's transmitter impairments, and $n(t)$ denotes additive white Gaussian noise (AWGN).

When RFFI is applied in a dynamic environment, the channel effect $h(t)$ is time-varying. Regarding receivers, it is highly likely that different receivers are used in the training and inference stages, resulting in different receiver impairments.
The channel effect $h(t)$ and the receiver effect $\mathcal{G}$ will change the data distribution of $x(t)$ both in the training and inference stages, therefore, mitigating the channel and receiver effect is essential. 
However, it is quite challenging to eliminate the effects in the time domain due to the convolutional process.

\subsection{Requirement of Labelled Training Data}\label{sec:shortage1}
In order to train a deep learning model with a good generalization capability, sufficient samples from each DUT are required in the training dataset, which significantly increases the collection overhead and labour costs. 
In the computer vision community, it is common to fine-tune a pretrained deep learning model using a small number of samples. However, in the RFFI community, due to the lack of a comprehensive dataset, many researchers tend to collect their own training datasets and train the RFFI model from scratch.

There are indeed a few RFFI datasets publicly available~\cite{shen2022towards, shen2023towards, lingnan2025infocom, al2021deeplora}. However, RFFI relies on the hardware impairments of DUTs, and it is unlikely the datasets would have the same DUTs. In other words, their label spaces are different. 

\section{System Overview}\label{sec:system}
\begin{figure}[!t]
	\begin{center}
		\includegraphics[width = 3.0in]{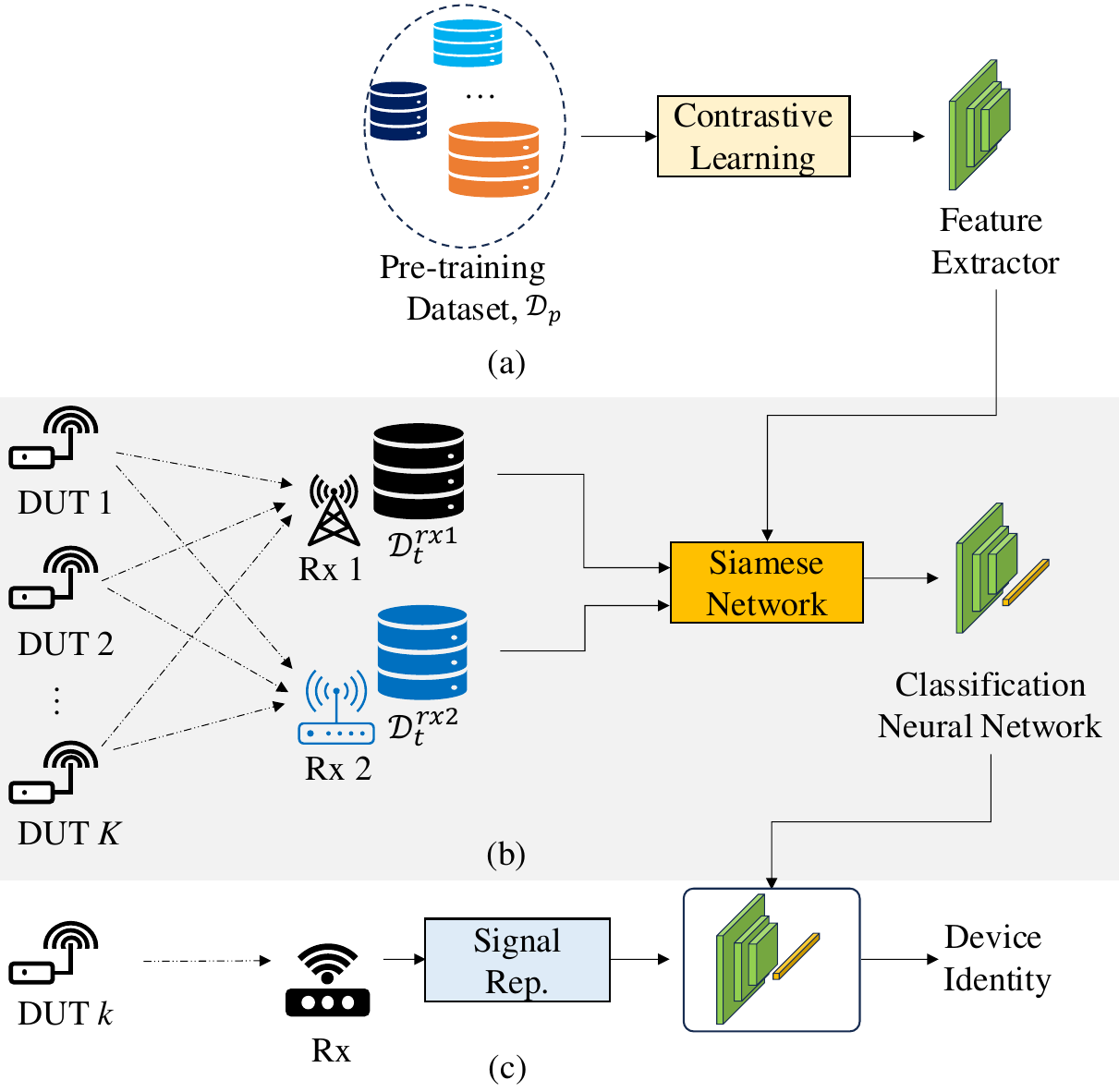}
		\caption{The proposed three-stage RFFI approach. (a) Contrastive learning-enhanced pretraining stage. (b) Siamese network-based training/fine-tuning stage. (c) Inference stage.}
		\label{fig:system}
	\end{center}	
\end{figure}

\mj{To address the above challenges, a three-stage RFFI approach is designed, including pretraining, fine-tuning, and inference, as shown in Fig.~\ref{fig:system}.} Specifically, unsupervised contrastive learning is adopted in the pretraining stage to obtain a channel-robust RFF feature extractor. In the fine-tuning stage, a Siamese network enhanced with contrastive loss is employed to eliminate the effects of channel and receiver variations. 

\subsection{Pretraining}\label{sec:pretraining}
As shown in Fig.~\ref{fig:system}(a), the pretraining stage used contrastive learning to train a feature extractor.
Firstly, the pretraining dataset, $\mathcal{D}_p$, can be created by combining public datasets. These datasets may involve different numbers of DUTs and receivers.
Unsupervised learning is adopted in this stage, which allows us to leverage the collected signal samples without worrying about the reliability of their labels.
Therefore, contrastive learning, which excels at capturing the representative features by treating the signal itself as the label, is employed. 
This approach is based on maximizing the similarity between two augmented versions of the same sample. 
The data augmentation here is used to simulate signals with various channel and noise effects, enabling the development of a channel-independent feature extractor.
After the pretraining is completed, a feature extractor $\mathcal{T}\left(\cdot, \theta_{\mathcal{T}}\right)$ will be returned, where $\theta_{\mathcal{T}}$ is the learnable parameters of the feature extractor. 
The details of pretraining will be presented in Section~\ref{sec:contrasRF}.

\subsection{Fine-Tuning}\label{sec:fine-tune}
\mj{As shown in Fig.~\ref{fig:system}(b), a Siamese network with two branches is adopted in the fine-tuning stage. While the Siamese network is usually designed for pairwise similarity comparison, this paper uses it differently for the multi-class classification task, which is achieved by using the cross-entropy loss together with the contrastive loss.

A neural network, $\psi\left(\cdot, \theta_{\psi}\right)$, will be created by concatenating the pretrained feature extractor and a dense layer, $\mathcal{C}\left(\cdot, \theta_{\mathcal{C}}\right)$, for classification, given as
\begin{equation}
\psi\left(\cdot, \theta_{\psi}\right)=\mathcal{T}\left(\cdot, \theta_{\mathcal{T}}\right) \circ \mathcal{C}\left(\cdot, \theta_{\mathcal{C}}\right),
\end{equation}
where $\theta_{\psi}$ and $\theta_{\mathcal{C}}$ represent the learnable parameters of the neural network and classification head, respectively. 

There are $K$ DUTs to be classified in this work.
To solve the receiver issue, we deliberately employ different receivers, with at least two receivers, to capture signals from these $K$ DUTs to construct fine-tuning datasets, i.e., $\mathcal{D}_t^{rx1}$ and $\mathcal{D}_t^{rx2}$.
We select packets from the same DUT but different receivers as a pair to send into the Siamese network. 
This approach facilitates obtaining receiver-independent features.
In addition, data augmentation will be used to enhance the neural network to be channel independent. 

After fine-tuning, a neural network $\psi\left(\cdot, \theta_{\psi}\right)$ that is receiver and channel independent will be obtained.
The fine-tuning details, such as deep learning architecture and loss function, will be introduced in Section~\ref{sec:siameseRFFI}.}

\subsection{Inference}\label{sec:inference}
\mj{Siamese networks usually involve two branches for pairwise similarity comparison during the inference stage. We use it different by only using one branch for multi-class classification, as illustrated in Fig.~\ref{fig:system}(c).

First, a receiver acquires a packet from the DUT that needs to be identified.
Second, the received signal is transformed into a representation, e.g., a spectrogram, which is then directly input into the neural network for classification.
Finally, the prediction of the signal will be given as 
\begin{equation}
    \hat{P} = \psi\left(X, \theta_{\psi}\right),
\end{equation}
where $X$ is the signal representation, and $\hat{P}$ is a label predicted by the model $\psi\left(\cdot, \theta_{\psi}\right)$ for the most likely DUT of the input.}

\section{Training/Fine-Tuning RFFI Using Siamese Network}\label{sec:siameseRFFI}
\begin{figure}[!t]
	\begin{center}
		\includegraphics[width = 3.4in]{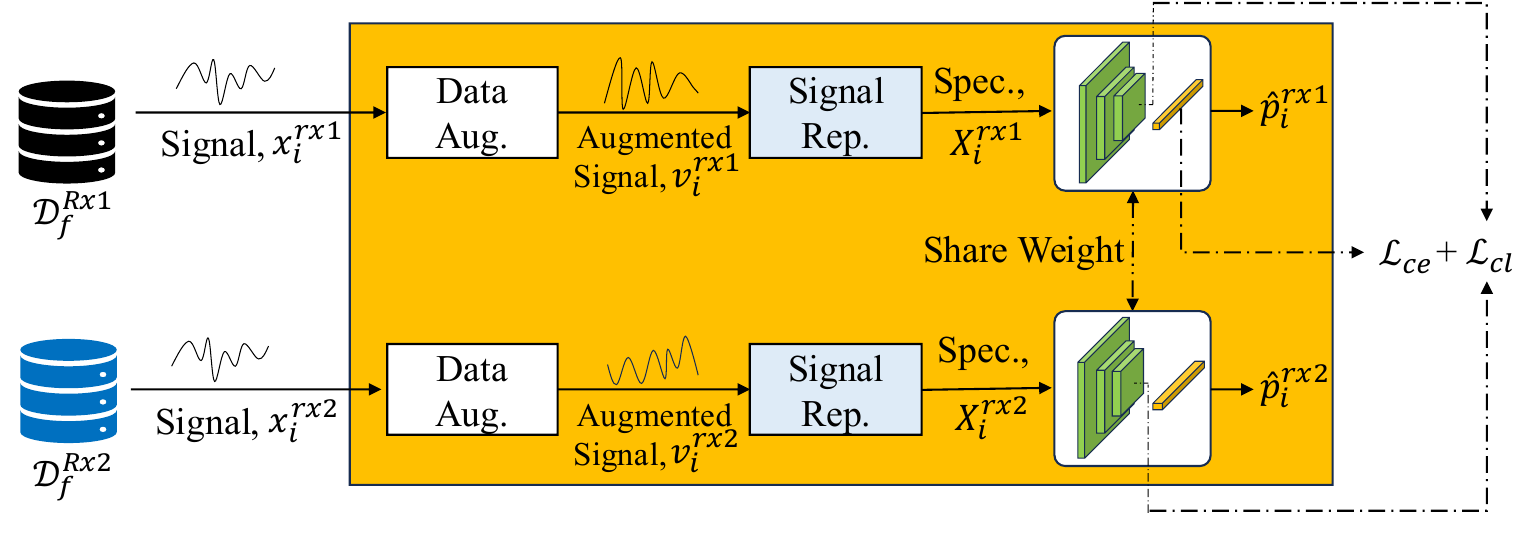}
		\caption{\mj{The proposed Siamese network-based training/fine-tuning process.}}
		\label{fig:siamese}
	\end{center}	
\end{figure}

The RFFI is sensitive to the variations of the wireless channel and the receivers. 
To address these challenges, we designed a Siamese network-based RFFI approach, as shown in Fig.~\ref{fig:siamese}. The model can be trained from scratch or fine-tuned from a pretrained model. The technical algorithm design of the above two schemes is the same.

\subsection{Overview}
We first constructed a training dataset, consisting of two sub training datasets. Each sub-dataset includes the time domain signals (IQ samples) from the same set of DUTs but captured by different receivers, i.e., $\mathcal{D}_t^{rx1}$ and $\mathcal{D}_t^{rx2}$. Every time, we selected a signal from each subdataset, $x_i^{rx1}$ and $x_i^{rx2}$, which have the same DUT label. 
In this paper, in order to reduce the overhead of collecting an abundant training dataset, we deliberately captured signals with high signal-to-noise ratio (SNR). Then, we used data augmentation to enrich the signals with channel variations and noise, which will be introduced in Section~\ref{sec:data_aug}.
Afterwards, the augmented signals will be transformed into a specific representation, namely spectrogram in this paper, in order to disentangle channel and receiver effects from the transmitter impairments, which will be presented in Section~\ref{sec:signalrep}.
The signal after representation is fed into the Siamese network model, which is composed of two identical subnetworks with shared weights and parameters. Enhanced by the contrastive loss, the Siamese network allows the model to effectively learn the similarities between input pairs through a common feature space, which will mitigate the channel and receiver effects. 
\mj{The overall loss function is formulated as the sum of a contrastive loss $\mathcal{L}_{\textit{cl}}$ and a cross-entropy loss $\mathcal{L}_{\textit{ce}}$. 
The detailed architecture of the Siamese network, along with the involved losses, is presented in Section~\ref{sec:siamese}.}

\subsection{Data Augmentation}\label{sec:data_aug}
Data augmentation serves as a crucial part of the system, as it enriches training datasets.
\mj{Both channel and AWGN augmentation are used. 
For the channel model, the tapped delay line (TDL) channel has been chosen.
The multipath effect is modelled using the exponential power delay profile (PDP), while the Doppler effect is characterized by the Jakes model.
Additionally, the synthetic AWGN is incorporated in data augmentation.
The multipath and AWGN augmentation of the system ensures the system is generalized and performs well under different scenarios. 
Regarding the ranges of the aforementioned terms, the RMS delay spread ranges from $[5, 300]$ ns, Doppler frequency spans $[0, 5]$ Hz, and SNR is $[10, 40]$ dB. 
These configurations are designed to expose the model to diverse channel conditions and SNR levels during training, thereby improving its robustness and generalization in real-world scenarios.}
Please refer to \cite{shen2022towards} for more information.

The simulation parameters mentioned above are randomized for each signal.
Online data augmentation~\cite{shen2023length} is utilized, which applies data augmentation during each epoch in training. It contributes to increasing the training dataset and improving the model's generalization.

After data augmentation, the original signals, ${x}_i^{rx1}$ and ${x}_i^{rx2}$, become ${v}_i^{rx1}$ and ${v}_i^{rx2}$, respectively. Because they are from the same DUT, they later form a positive pair for the Siamese network.
\mj{It is important to note that each sample in a training batch is independently augmented twice to form a positive pair. 
Moreover, random augmentation is applied at each training epoch, thereby exposing each signal instance to a diverse set of channel and noise conditions over time.
This variability enables the model to learn robust and transmitter-specific representations that generalize effectively across a broad range of propagation environments.}

\subsection{Siganl Representation}\label{sec:signalrep}
Signal representation determines the neural network's input, which could be time-domain, frequency-domain, or other forms.
In this paper, we employed the spectrogram, which reveals the time-frequency domain features, in order to separate the transmitter impairments from the receiver impairments and channel effects.

The time-domain IQ signal, $x(t)$ in (\ref{eq:time_signal}), can be transformed to the spectrogram in time-frequency domain, $X(t, f)$, by STFT, given as
\begin{equation}
\begin{aligned}
X(t, f) & = {\mathcal{S}}\big(x(t)\big) \\
& = \mathcal{S}\big(\mathcal{G}\big) \mathcal{S}\big(h(t)\big) \mathcal{S}\big(\mathcal{F}_k(s(t))\big) + \mathcal{S}\big(n(t)\big),
\end{aligned}
\end{equation}
where ${\mathcal{S}}(\cdot)$ represents STFT operation.
We then convert the amplitude of the spectrogram from the linear scale to the logarithmic scale, which can be expressed as
\begin{align}
&\log |X(t, f)|  =\log|\mathcal{S}\big (\mathcal{G}\big) \mathcal{S}\big(h(t)\big) \mathcal{S}\big(\mathcal{F}_k(s(t))\big) + \mathcal{S}\big(n(t)\big)|\nonumber\\
& \approx \log |{\mathcal{S}}(\mathcal{G})| + \log |{\mathcal{S}}(h(t))| +\log |{\mathcal{S}}(\mathcal{F}_k(s(t))|.
\label{eq:spec}  
\end{align}
\mj{The approximation holds when the signal has a high SNR level.}

As shown in (\ref{eq:time_signal}), the time domain signal involves both channel and receiver effects, which are difficult to separate due to the convolution operation. After it is converted to the spectrogram, the effects from the channel and receiver still exist. However, they become summation terms in the spectrogram, which are easier to separate. 

After signal representation, $x_i^{rx1}$ and $x_i^{rx2}$ can be represented as follows:
\begin{align}
x_i^{rx1} \rightarrow \textcolor[rgb]{1,0,0}{\log |{\mathcal{S}}(\mathcal{G}_1)|+ \log |{\mathcal{S}}(h_1(t))|}+\log |{\mathcal{S}}(\mathcal{F}_k(s(t))|, \label{eq:spec1}\\
x_i^{rx2} \rightarrow \textcolor[rgb]{1,0,0}{\log |{\mathcal{S}}(\mathcal{G}_2)|+ \log |{\mathcal{S}}(h_2(t))|}+\log |{\mathcal{S}}(\mathcal{F}_k(s(t))|,\label{eq:spec2}
\end{align}
where the terms highlighted in red are different between two signals, and the last term is common when $x_i^{rx1}$ and $x_i^{rx2}$ are from the same DUT.

\subsection{Siamese Neural Network}\label{sec:siamese}
As shown in Fig.~\ref{fig:siamese}, different from a common deep learning network, a Siamese neural network involves two branches with the same architecture and shared weights. 

\subsubsection{Deep Learning Architecture}\label{sec:dlarchitecture}
\begin{figure}[!t]
	\begin{center}
		\includegraphics[width = 3.4in]{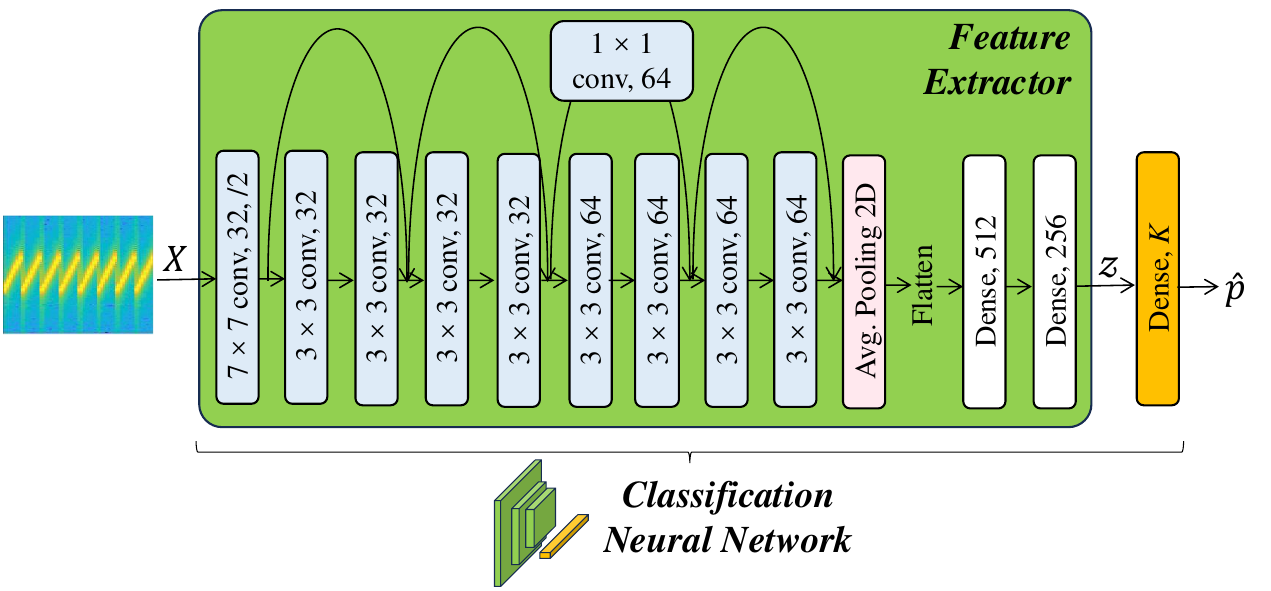}
		\caption{The architecture of ResNet-based deep learning model consisting of a feature extractor and classifier.}
		\label{fig:resnet}
	\end{center}	
\end{figure}

The deep learning model is portrayed in Fig.~\ref{fig:resnet}. 
As spectrograms can be regarded as 2D images, we designed the model based on ResNet, which is a popular convolutional neural network (CNN) architecture.

A classification CNN model can be further partitioned into a feature extractor part and a classifier. 
\begin{itemize}
	\item The \textit{feature extractor} consists of nine convolutional layers, one average pooling layer, and two dense layers. 
The first convolutional layer uses 32 $7 \times 7$ kernels. The second to the fifth layers use 32 $3 \times 3$ kernels, while the sixth to the ninth layers employ 64 $3 \times 3$ kernels. 
Skip connections are employed to avoid gradient vanishing and facilitate optimization.
The fully connected layers consist of $512$ and $256$ neurons, respectively. The output vector, $z$, serves as the RFF feature. 
	\item The classifier consists of one dense layer with $K$ neurons. It will make the prediction based on the RFF feature.
\end{itemize}

\subsubsection{Loss Function}\label{sec:calculate_loss}
The Siamese network is optimized by the contrastive loss, $\mathcal{L}_{\textit {cl}}$, and cross-entropy, $\mathcal{L}_{\textit {ce}}$.
The loss function $\mathcal{L}$ is given as
\begin{equation}
\mathcal{L} = \mathcal{L}_{\textit {cl}} + \mathcal{L}_{\textit {ce}}.
\end{equation}
The loss $\mathcal{L}_{\textit {cl}}$ and $\mathcal{L}_{\textit {ce}}$ is calculated by summing $\ell_{\textit {i, j}}$ and $\ell_{\textit {ce}}$ within a batch $\mathcal{B}$, respectively.

\textbf{Contrastive loss} plays a pivotal role in building the receiver and channel independent RFFI system.
It enables the Siamese network to learn the common transmitter impairments and mitigate other effects between the samples in the two branches, which are from the same DUT but received by different receivers and augmented through distinct channels.

Contrastive loss maximizes the similarity of the positive pairs.
A comparison between (\ref{eq:spec1}) and (\ref{eq:spec2}) reveals that the term $(\log |{\mathcal{S}}(\mathcal{F}_k(s(t))|)$ remains identical in both expressions for the positive pair, which represents the intrinsic characteristics of the device.
Meanwhile, the effect of $(\log |{\mathcal{S}(\mathcal{G}_1)}| + \log |{\mathcal{S}}(h_1(t))|)$ and $(\log |{\mathcal{S}(\mathcal{G}_2)}| + \log |{\mathcal{S}}(h_2(t))|)$, representing variations introduced by receivers and channels, will be mitigated. 
This ensures that the model learns the RFFs solely, without being influenced by other factors.

NT-Xent loss~\cite{chen2020simple} is an effective and widely used contrastive loss, which was applied in our work. 
Given a set $\left\{{{x}}_k\right\}$ including a positive pair of examples ${x}_i$  and ${x}_j$, the contrastive loss helps to identify ${x}_j$ in $\left\{{x}_k\right\}_{k \neq i}$ for a given ${x}_i$.
The loss function for each positive pair $(i, j)$ is defined as 
\begin{equation}\label{eqn:sim}
\ell_{i, j}=-\log \frac{\exp \left(\operatorname{sim}\left({z}_i, {z}_j\right) / \tau\right)}{\sum_{k=1}^{2 |\mathcal{B}|} \mathds{1}_{[k \neq i]} \exp \left(\operatorname{sim}\left({z}_i, {z}_k\right) / \tau\right)},
\end{equation}
where $\operatorname{sim}({a}, {b})={a}^{\top} {b} /\|{a}\|\|{b}\|$ denotes the dot product between $l_2$ normalized ${a}$ and ${b}$ (i.e., cosine similarity). $\mathds{1}_{[k \neq i]} \in\{0,1\}$ is an indicator function that evaluates to $1$ if $k \neq i$, $\tau$ denotes a temperature parameter (set as 0.05 in our work), and $|\mathcal{B}|$ is the batch size. 
The similarity between RFF ${z}_i$ and ${z}_j$ will be maximized by (\ref{eqn:sim}), and the total loss $\mathcal{L}_{\textit {cl}}$ is computed across all positive pairs in a batch $\mathcal{B}$. 
\mj{In our implementation, $|\mathcal{B}|$ is 32, each signal generates two augmented views, resulting in a total of 64 samples. 
Each sample serves as an anchor, with its counterpart forming the positive pair, and the remaining 62 samples acting as negatives for contrastive loss computation. All augmented samples are used as anchors in turn, thereby fully utilizing the pairwise relationships within each batch.}

\textbf{Cross-entropy} assists the neural network in classification. 
The calculation is mathematically given as
\begin{equation}
\ell_{ce}=-\sum_{k=1}^{K} q_k \log \left(\hat{q}_k\right),
\end{equation}
where $q_k$ is the ground truth, and $\hat{q}_k$ is the probabilities of the $k^{th}$ label, returned by the softmax activation function.




\section{Pretraining Feature Extractor Using Contrastive Learning}\label{sec:contrasRF}
While the deep learning approach presented in Section~\ref{sec:siameseRFFI} can work standalone, it indeed requires a large amount of labelled training dataset.
It is common to pretrain the feature extractor $\mathcal{T}\left(\cdot, \theta_{\mathcal{T}}\right)$ and then use a few labelled samples to fine-tune the neural network.

\begin{figure}[!t]
	\begin{center}
		\includegraphics[width = 3.4in]{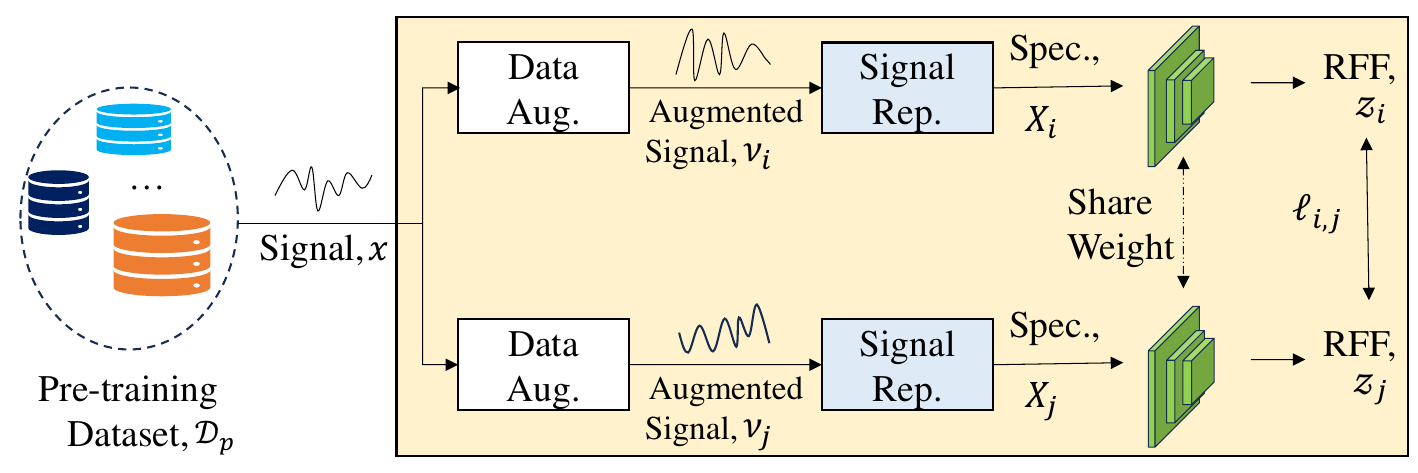}
		\caption{Pretraining the feature extractor with contrastive learning.}
		\label{fig:contasRFF}
	\end{center}	
\end{figure}

The process of obtaining the feature extractor with contrastive learning in the pretraining stage is shown in Fig.~\ref{fig:contasRFF}. 
Although it has the same data augmentation and signal representation as in Fig.~\ref{fig:siamese}, the design philosophy is totally different.
Specifically, the pretraining is done via unsupervised learning, while the training/fine-tuning in Section~\ref{sec:siameseRFFI} works in a supervised learning manner.

The pretraining dataset can be from a single source or from multiple sources. While the label information may not be reliable, we only use the signal samples, hence unsupervised learning is required.

A signal $x$, will be randomly selected from the pretraining dataset. It then will be individually augmented by adding random channel interferences and noise, via data augmentation. 
The process and parameter settings of the data augmentation are the same as Section~\ref{sec:data_aug}. After data augmentation, each signal will become two copies, ${v}_1$ and ${v}_2$.

After that, both copies are converted to the same signal representation used in Section~\ref{sec:signalrep}, i.e., the spectrogram in this paper. The two signals then become
\begin{align}
v_{1} \rightarrow \log |{\mathcal{S}}(\mathcal{G})|+ \textcolor[rgb]{1,0,0}{\log |{\mathcal{S}}(h_1(t))|}+\log |{\mathcal{S}}(\mathcal{F}_k(s(t))|, \label{eq:specx1}\\
v_{2} \rightarrow \log |{\mathcal{S}}(\mathcal{G})|+ \textcolor[rgb]{1,0,0}{\log |{\mathcal{S}}(h_2(t))|}+\log |{\mathcal{S}}(\mathcal{F}_k(s(t))|.\label{eq:specx2}
\end{align}
Because the signals are augmented from the same signal, $x$, only the channel parts are different between them.

Finally, the spectrogram is used to train the feature extractor.
Because unsupervised learning is used, the loss function used in this stage is also the NT-Xent contrastive loss, which has been described in detail in Section~\ref{sec:calculate_loss}.

Thanks to contrastive learning, it can alleviate effects from channels and obtain device-intrinsic features, which eventually produce a channel-independent feature extractor.

\section{Experimental Setup and Datasets}\label{sec:setup}
In order to evaluate the proposed approach comprehensively against various channel and receiver effects, we deliberately chose two public LoRa datasets~\cite{shen2022towards,shen2023towards} and created our own dataset.
The evaluations are summarized in Table~\ref{sec:evaluation}.
\mj{It is noted that in this work, the training dataset consists of high-SNR samples with little channel and noise effects.
Data augmentation applies various channels and SNRs to these high SNR samples, allowing us to simulate diverse conditions.}
\begin{table*}[]
\caption{Summary of Experimental Evaluation}
\label{sec:evaluation}
\begin{tabular}{|l|l|l|L{1.44cm}|L{2.6cm}|L{5.34cm}|}
\hline
Section                        & Evaluation   Purpose                                              & Pretrain                  & Dataset                                                                                     & Training   Dataset                                                                         & Test Dataset                                                                                                           \\ \hline
VII-A                          & Channel elimination                                               & N                         &                                                                                             &                                                                                            &                                                                                                                        \\ \cline{1-3}
\cellcolor[HTML]{D9D9D9}VIII-B & \cellcolor[HTML]{D9D9D9}Pre-training, when   only channel effect  & \cellcolor[HTML]{D9D9D9}Y & \multirow{-2}{*}{\textit{\begin{tabular}[c]{@{}l@{}}Public\\ Dataset 1\end{tabular}}}       & \multirow{-2}{*}{\begin{tabular}[c]{@{}l@{}}10 DUTs\\ 1 Rx (N210)\end{tabular}}            & \multirow{-2}{*}{\begin{tabular}[c]{@{}l@{}}Channel: LOS \& NLOS and static \& dynamic \\ channels.\end{tabular}}      \\ \hline
VII-A                          & Channel elimination                                               & N                         & \textit{\begin{tabular}[c]{@{}l@{}}Public\\ Dataset 2\end{tabular}}                         & \begin{tabular}[c]{@{}l@{}}10 DUTs \\ 1 Rx (B210)\end{tabular}                             & Channel: Room \& Outdoor \&   Office.                                                                                  \\ \hline
VII-B                          & Receiver mitigation                                               & N                         &                                                                                             &                                                                                            &                                                                                                                        \\ \cline{1-3}
\cellcolor[HTML]{D9D9D9}VIII-C & \cellcolor[HTML]{D9D9D9}Pre-training, when   only receiver effect & \cellcolor[HTML]{D9D9D9}Y & \multirow{-2}{*}{\textit{\begin{tabular}[c]{@{}l@{}}Public\\ Dataset 3\end{tabular}}}       & \multirow{-2}{*}{\begin{tabular}[c]{@{}l@{}}10 DUTs \\ 2 Rx (B200-1, B210-1)\end{tabular}} & \multirow{-2}{*}{\begin{tabular}[c]{@{}l@{}}Channel: LOS, high SNR. \\ Rx: 20\end{tabular}}                            \\ \hline
VIII-C                         & Joint channel and   receiver mitigation                           & N                         &                                                                                             &                                                                                            &                                                                                                                        \\ \cline{1-3}
\cellcolor[HTML]{D9D9D9}VIII-D & \cellcolor[HTML]{D9D9D9}Joint channel and   receiver mitigation   & \cellcolor[HTML]{D9D9D9}Y & \multirow{-2}{*}{\textit{\begin{tabular}[c]{@{}l@{}}Public\\ Dataset 3\end{tabular}}}       & \multirow{-2}{*}{\begin{tabular}[c]{@{}l@{}}10 DUTs \\ 2 Rx (B200-1, B210-1)\end{tabular}} & \multirow{-2}{*}{\begin{tabular}[c]{@{}l@{}}Channel: NLOS and static (six locations). \\ Rx: 3 USRP N210\end{tabular}} \\ \hline
VII-D                          & Joint channel and   receiver mitigation                           & N                         &                                                                                             &                                                                                            &                                                                                                                        \\ \cline{1-3}
\cellcolor[HTML]{D9D9D9}VIII-E & \cellcolor[HTML]{D9D9D9}Joint channel and   receiver mitigation   & \cellcolor[HTML]{D9D9D9}Y & \multirow{-2}{*}{\textit{\begin{tabular}[c]{@{}l@{}}Self-Collected\\ Dataset\end{tabular}}} & \multirow{-2}{*}{\begin{tabular}[c]{@{}l@{}}10 DUTs\\ 2 Rx (B210-a, N210-c)\end{tabular}}  & \multirow{-2}{*}{\begin{tabular}[c]{@{}l@{}}Channel: LOS \& NLOS and dynamic channels.   \\ Rx: 6\end{tabular}}        \\ \hline
\end{tabular}
\end{table*}

\subsection{Datasets}\label{sec:dataset}
\subsubsection{Device Configuration}
All the datasets used in this paper follow the same device configurations. 
The LoRa DUTs were configured with bandwidth $B = 125$~kHz.
The sampling rate of the receiver SDR was $f_s = 1$~MHz. 
The Matlab Communications Toolbox Support Package for SDR was used for accessing IQ samples from SDR platforms\footnote{https://uk.mathworks.com/help/comm/supported-hardware-software-defined-radio.html}. Eight preambles at the start of each LoRa packet were captured, as exemplified in Fig.~\ref*{fig:waveform}(a).

\subsubsection{Public Dataset 1}\label{sec:publicdata1}
The work in~\cite{shen2022towards} released a LoRa RFFI dataset, which includes test datasets collected in different locations. 
The DUTs were ten LoPy4 devices, and the receiver was a USRP N210 SDR platform. 

\textbf{Training dataset:} There were 400 LoRa packets for each DUT, which were collected in a residential room with line-of-sight (LOS)  available. The DUT and the receiver were kept close (about half a meter) to obtain high SNR data.

\begin{figure}[!t]
\centering
\subfloat[]{\includegraphics[width=3.0in]{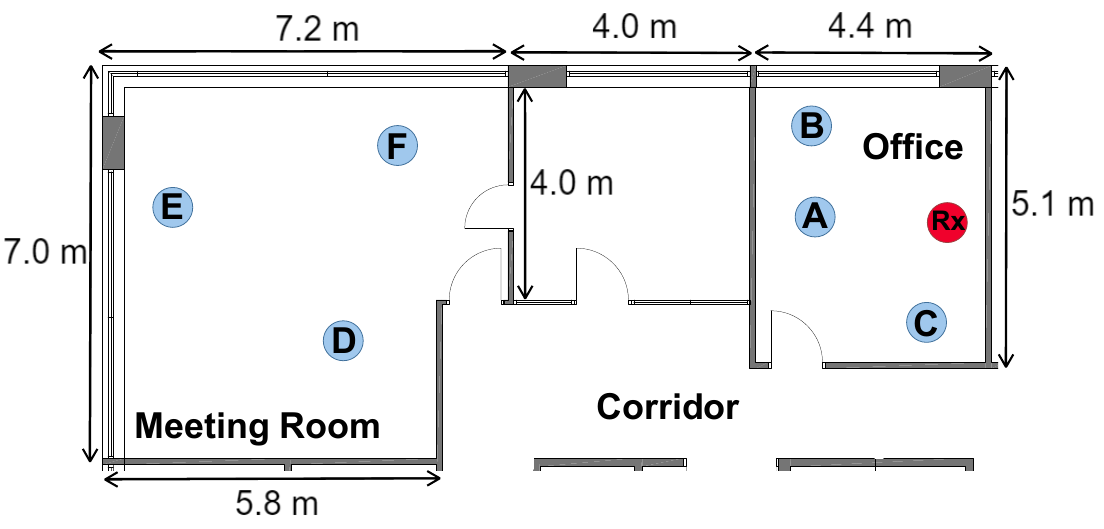}}

\subfloat[]{\includegraphics[width=3.4in]{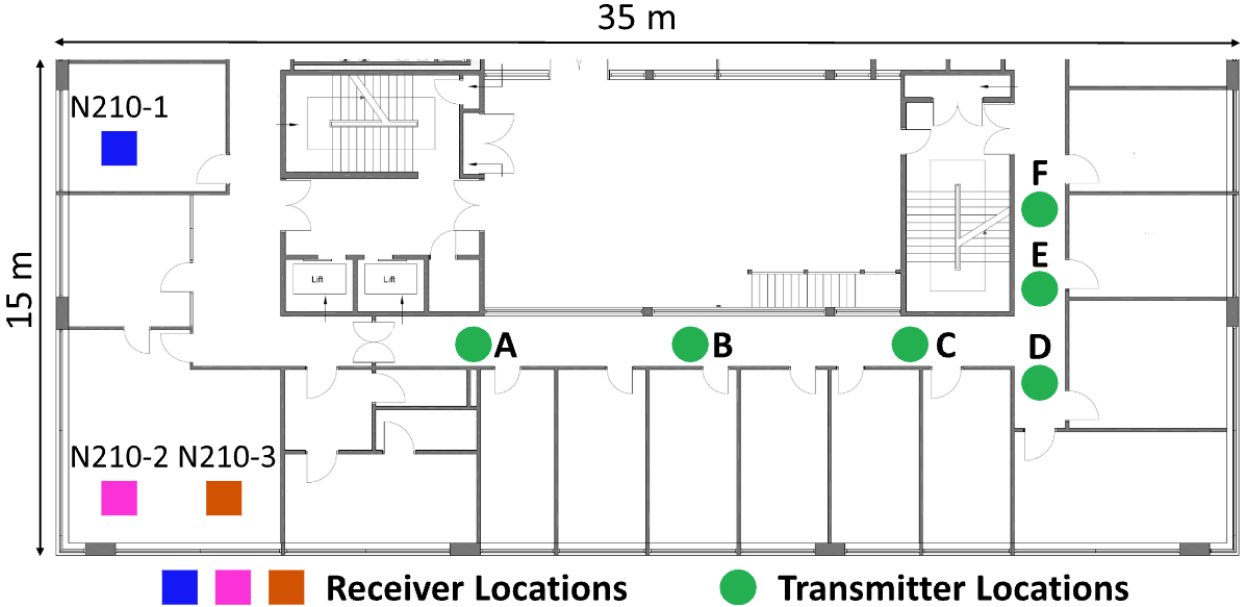}
}

\subfloat[]{\includegraphics[width=3.4in]{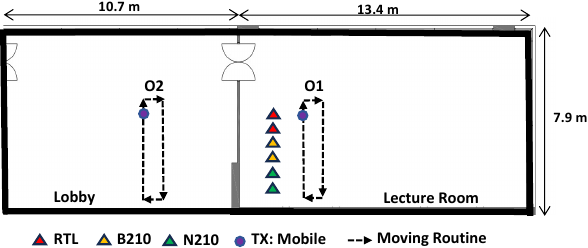}
}
\centering
\caption{Floor plan of the experimental environments. (a) \textit{Public Dataset 1}~\cite{shen2022towards}. (b) \textit{Public Dataset 3}~\cite{shen2023towards}. (c) Self-collected dataset.}
\label{fig:floorplan}
\end{figure}
\textbf{Test dataset:}
There were 200 packets per DUT. There were both static and dynamic channel scenarios, as well as LOS and non-line-of-sight (NLOS). The floorplan of the dataset collection position is shown in Fig.~\ref{fig:floorplan}(a). The USRP N210 SDR platform was always placed at the red Rx position.
\begin{itemize}
    \item Static channel: Test datasets D1-D6 were collected at six locations, labelled as A-F, respectively. 
    The DUTs remained stationary, with no moving objects nearby. 
    \item Dynamic channel: There were two cases. (i) Objects moving: D7 and D8 were collected at Location B and F, respectively, with the DUTs kept stationary while a person walked randomly around the room at a speed of 2 m/s. (ii) Mobile device: D9 and D10 were respectively collected in an office and meeting room, where a person walked around carrying the DUTs at a speed of 2 m/s.
\end{itemize}
Datasets D1, D2, D3, D7, and D9 were collected in  LOS environments, while D4, D5, D6, D8, and D10 were collected in  NLOS environments.

\subsubsection{Public Dataset 2}\label{sec:public3}
\mj{The work in~\cite{2021access} released a LoRa RFFI dataset, which includes data collected in different locations, utilizing USRP B210 as the receiver and Pycom as the transmitters. 
For our evaluation, Dev 1 through Dev 10 were employed, and the preamble portion of each signal was extracted and utilized for RFF.

\textbf{Training dataset:} There were 400 LoRa packets for each DUT.
All devices transmit an identical message from a fixed location, positioned 5 meters from the receiver, ensuring that each device experienced similar channel conditions. Data collection was conducted in an occupied indoor room.

\textbf{Test dataset:} 
There were 40 packets per DUT. 
LoRa transmissions were collected from three distinct deployment environments, i.e., an indoor room (Loc.~1), an office space (Loc.~2), and an outdoor area (Loc.~3), all on the same day.}

\subsubsection{Public Dataset 3}\label{sec:publicdata2_2}
The work in~\cite{shen2023towards} shared a LoRa RFFI dataset related to various receivers.
The DUTs were ten LoRa devices, including five LoPy4 and five mbed 1261 shields. Various receiver SDR platforms were used.

\textbf{Training dataset:} 
$800$ LoRa packets for each DUT-receiver pair were collected by B200-1 and B210-1, at an approximate distance of half a meter. The signals were with high SNR.

\textbf{Test dataset A:}
For the receiver, twenty SDR platforms of six types were used, including nine RTL, two Pluto, two B200, two B200 mini, two B210, and three N210.
There were 200 packets per DUT. All the devices remained stationary during data collection, located half a meter apart from the receiver. The signals were with high SNR.

\textbf{Test dataset B:}
This test dataset involved different receivers and static channel effects.
The involved receivers were N210-1/2/3 and the DUTs were placed in various locations.
The floor plan of the experimental environment is given in Fig.~\ref{fig:floorplan}(b).
N210-1 was positioned in an office, while N210-2 and N210-3 were placed in a meeting room. 
The LoRa DUTs were tested at six locations, labelled as A to F. The SNR across Locations A to F decreased from $50$~dB to $10$~dB.

\subsubsection{Self-Collected Dataset}\label{sec:selfdata}
\begin{figure}[!t]
\centering
\subfloat[]{\includegraphics[width=1.6in]{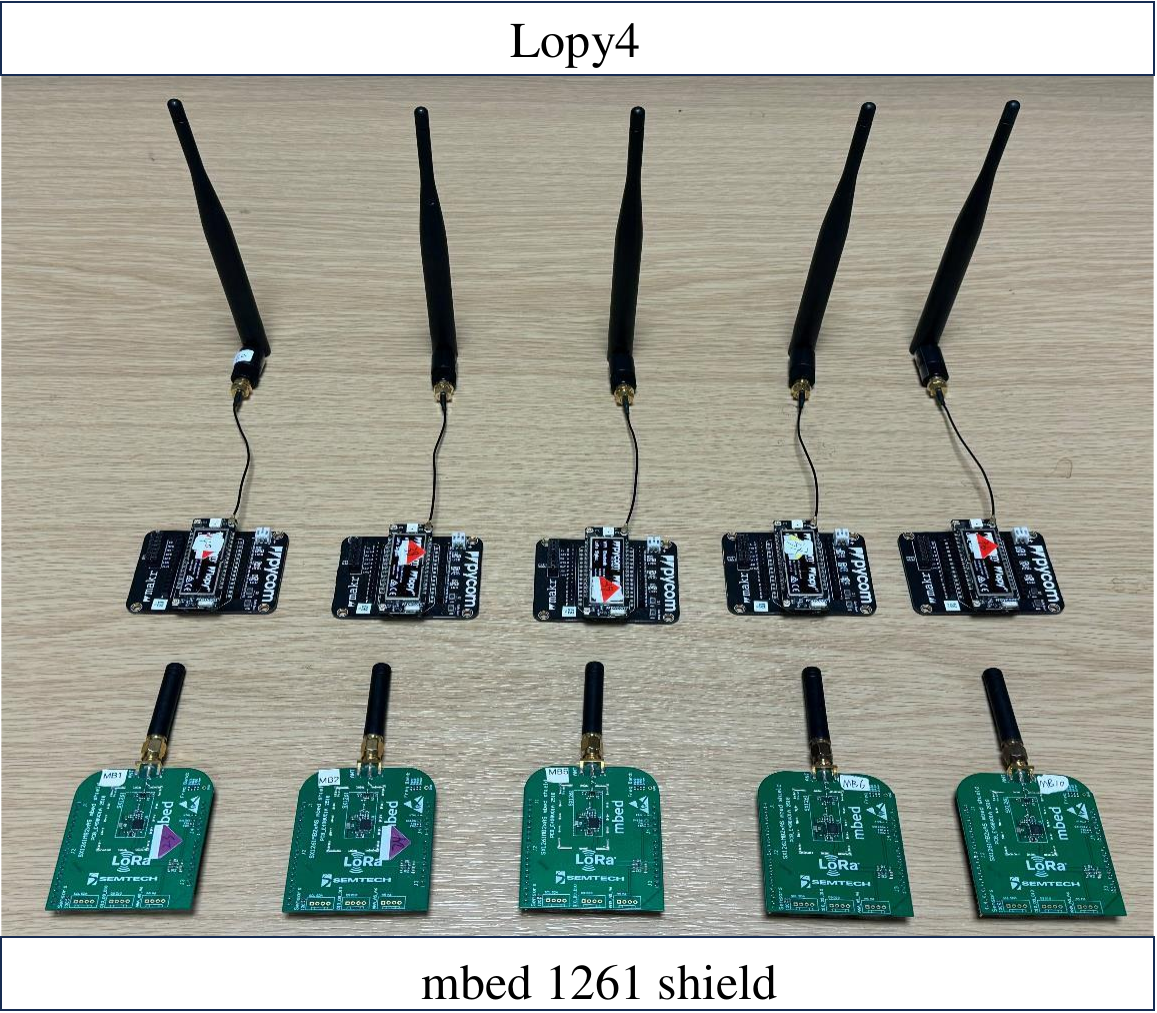}\label{fig:lora}}
\hspace{0.4cm}
\subfloat[]{\includegraphics[width=1.6in]{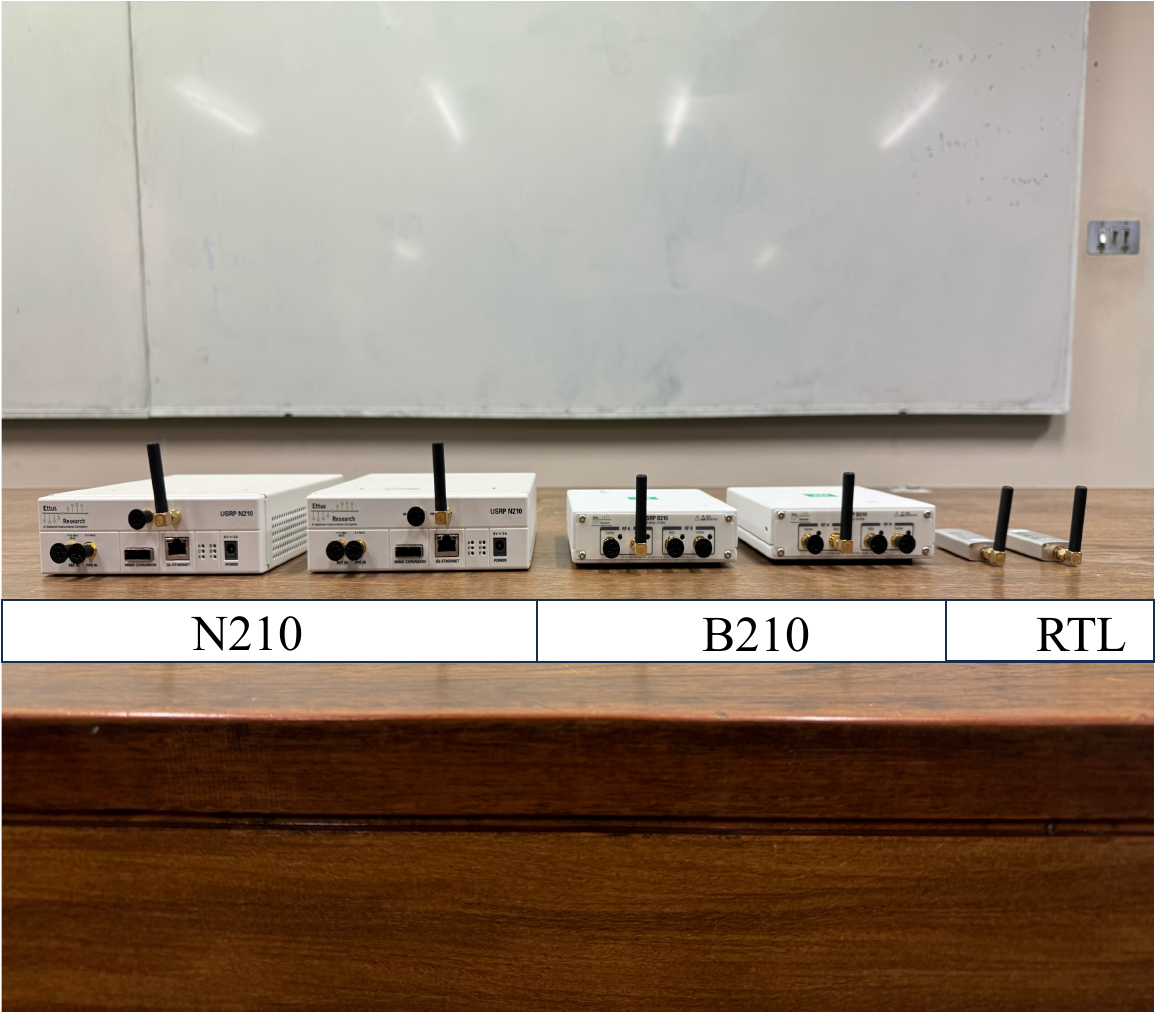}\label{fig:usrp}}
\caption{Experimental devices. (a) DUT: ten LoRa devices. (b) Receiver: six SDR receivers. }
\label{fig:equip}
\end{figure}

The work in~\cite{shen2023towards} only involved static channel effects, therefore, we created a dataset involving both different receivers and dynamic channel effects.
As shown in Fig.~\ref{fig:equip}(a), five LoPy4 and five mbed 1261 shields were used as DUTs. 

\textbf{Training dataset:} There were 800 LoRa packets for each DUT-receiver pair. 
During the data collection, the DUT and the receiver were placed stationary, one meter apart, with a LOS between them. 
The receivers used were B210-a and N210-c. where N210-c was not used in the test datasets.

\textbf{Test dataset:} 
Fig.~\ref{fig:floorplan}(c) displays the floor plan of the environment where the test datasets were collected.
The receivers involved consisted of two N210, two B210, and two RTL, labelled as N210-a/b, B210-a/b, and RTL-a/b as illustrated in Fig.~\ref{fig:equip}(b).
The test dataset includes 200 packets per DUT-receiver pair. 
The receiver was kept stationary while a person held a DUT and walked along routes O1 and O2 at a speed of 2 m/s for LOS and NLOS mobile scenarios, respectively.

\subsection{Benchmark Approaches}
We employed the following benchmark algorithms for comparing the effects of channel mitigation, with different signal representations, e.g., channel independent spectrogram (CIS), spectrogram, and IQ samples, which are introduced as below.
\begin{itemize}
    \item \mj{\textit{CIS}~\cite{shen2022towards}: This state-of-the-art method was designed to mitigate channel effects while preserving device-specific characteristics for reliable RFF.
    The representation is shown in Fig.~\ref*{fig:waveform}(c).}
    \item  \mj{\textit{CIS + Siamese}: CIS was used as the signal representation, and the deep learning model was trained with the Siamese network.}
    \item \mj{\textit{Spec}: Spectrogram was used as the signal representation.}
    \item \textit{IQ + Siamese}: IQ sample was as the signal representation, and the deep learning model was trained with the Siamese network. 
    \item \textit{IQ}: IQ sample was used as the signal representation.  
\end{itemize}
For \textit{CIS}, \textit{Spec} and \textit{IQ}, the deep learning was trained in the conventional way, i.e., the Siamese network was not involved.
For \textit{IQ}, the network architecture was also adapted from ResNet, as Fig.~\ref{fig:resnet} shows, but the kernel size was adjusted according to the dimensions of the IQ samples.

When there were different receivers used, another two benchmark algorithms were also used.
\begin{itemize}
    \item \mj{ \textit{CIS + Reverse} \cite{shen2023towards}: This state-of-the-art method adopted CIS as the signal representation and employed a gradient reversal layer to mitigate the influence of the receiver.}
    \item \textit{Spec + Reverse}: The spectrogram was used as the signal representation, and the gradient reversal layer was applied.
\end{itemize}
For the above two approaches, the Siamese network was not involved.

\subsection{Training Configuration}\label{sec:traincon}
The deep learning training was carried out on a PC equipped with a GPU of NVIDIA GeForce RTX 4090, and PyTorch was used. The neural network parameters were optimized by the adaptive moment estimation (Adam) optimizer with an initial learning rate of 0.0003 and a batch size of 32. The validation loss was monitored during training, and the learning rate was reduced by a factor of 0.5 when the validation loss did not drop within 10 epochs. Training terminated when the validation loss did not change within 30 epochs.

\subsection{Metric}
The classification accuracy is the primary evaluation criterion for the RFFI system, expressed as
$$
\text { Accuracy }=\frac{\text { Number of correctly classified packets }}{\text { Total number of packets }} \text {. }
$$

\section{Experimental Results of Channel and Receiver Mitigation} \label{sec:eva_siamese}
In this section, we conducted a comprehensive evaluation of the Siamese network's effectiveness in eliminating the influence of channel and receiver effects. There is no pretraining involved in this evaluation. Instead, we trained the Siamese network from scratch using the training datasets described in Section~\ref{sec:dataset}, which allowed us to study the effectiveness of Siamese network. In addition, we carried out ablation studies, by first investigating channel effect  (Section~\ref{sec:eva_channel}) and receiver effect  (Section~\ref{sec:eva_rx}) separately, and then combining both the channel and receiver effects (Sections~\ref{sec:eva_rx_channel_static} and~\ref{sec:eva_rx_channel_dynamic}).

\subsection{Evaluation Against Channel Effect}\label{sec:eva_channel}
\mj{The \textit{Public Dataset 1} introduced in Section~\ref{sec:publicdata1} only involved the channel effects because the same receiver was used for the training and test datasets. 
\begin{figure*}[!t]
\centering
\subfloat[]{\includegraphics[width=4in]{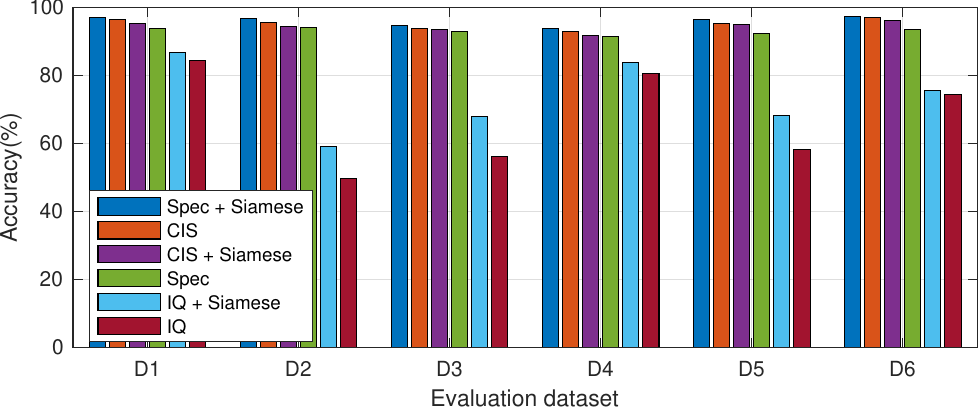}\label{fig:16}} \hspace{0.01in}
\subfloat[]{\includegraphics[width=1.5in]{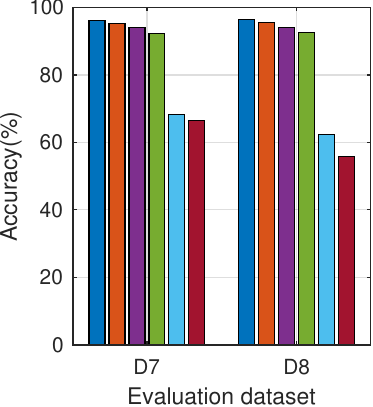}\label{fig:17}}
\hspace{0.01in}
\subfloat[]{\includegraphics[width=1.5in]{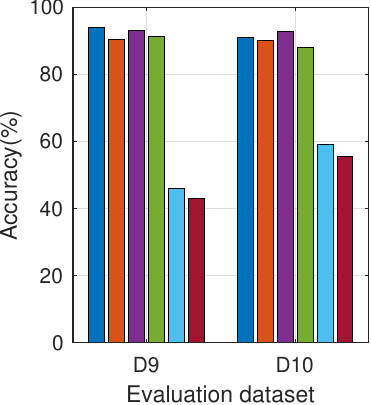}\label{fig:18}}
\centering
\caption{Evaluation of the proposed Siamese network-based RFFI for the channel effect mitigation. \textit{Public Dataset 1}~\cite{shen2022towards} was used. (a) Test datasets D1-D6 were collected in static scenarios. (b) D7 and D8 were collected in dynamic scenarios involving moving objects. (c) D9 and D10 were collected in dynamic scenarios involving the moving DUT.}
\label{fig:channels}
\end{figure*}

The classification results in various channels are presented in Fig.~\ref{fig:channels}.
\textit{Spec + Siamese} demonstrates the best performance across all test datasets.
In static scenarios, the accuracy of the \textit{Spec + Siamese} remains above 94\% in LOS scenarios and 93\% in NLOS scenarios. 
As for dynamic NLOS scenarios, its accuracy is still above 91\%, which indicates that the proposed \textit{Spec + Siamese} approach exhibits robustness across diverse channel conditions.
When IQ samples were used, their performance was always worse than the spectrogram-based approaches, because the channel effects are convolved with the signal in the time domain and more difficult to eliminate. 

The performance under dynamic channel conditions (D7 to D10) is worse compared to those under the static channels (D1 to D6), which is attributed to the increased multipath and Doppler effects in dynamic channels. The trend matched the results in~\cite{shen2022towards}. It is worth mentioning that our \textit{Spec + Siamese} has an accuracy of 91.2\% in the most challenging D10 scenario. Furthermore, our \textit{Spec + Siamese} approach always outperformed the \textit{CIS} and \textit{CIS + Siamese}.
It is worth noting that both CIS and Siamese network designs aim to reduce the impact of channel effects.
To perform a more comprehensive comparison, we additionally included the approach, \textit{CIS + Siamese}, that combines these two methods.
However, its overall performance is inferior to that of \textit{Spec + Siamese}. 
This may be attributed to the use of CIS as the feature representation, since the division operations required in its computation could diminish hardware-relevant features.
\begin{figure}
\centerline{\includegraphics[width=3.4in]{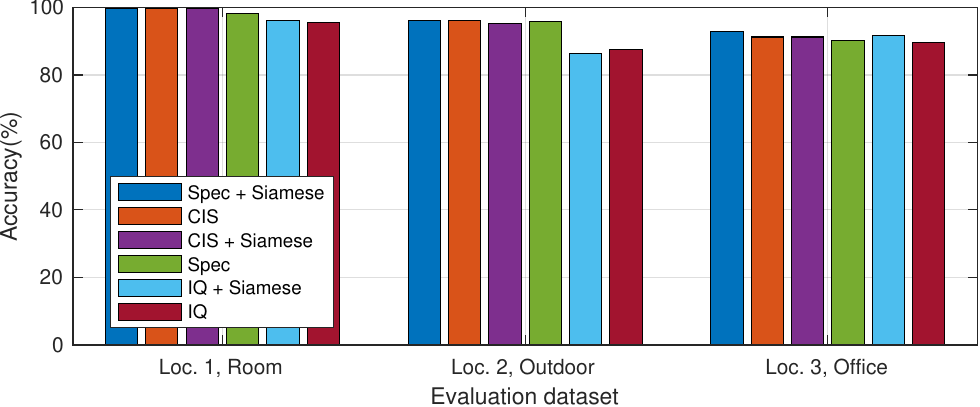}}
\caption{Evaluation of the proposed Siamese network-based RFFI for the channel effect mitigation with the dataset published by \textit{Public Dataset 2}~\cite{2021access}.}
\label{fig:orgen}
\end{figure}}

\mj{To further validate the effectiveness of the proposed method, we conducted experiments on \textit{Public Dataset 2}, and the corresponding results are shown in Fig.~\ref{fig:orgen}.
The classification trends remained consistent with the former results, the proposed \textit{Spec + Siamese} method achieved the best overall performance across all scenarios. 
In addition, IQ-based methods were observed to be less competitive.}

\subsection{Evaluation Against Receiver Effect}\label{sec:eva_rx}
\begin{figure*}[!t]
	\begin{center}
		\includegraphics[width = 7in]{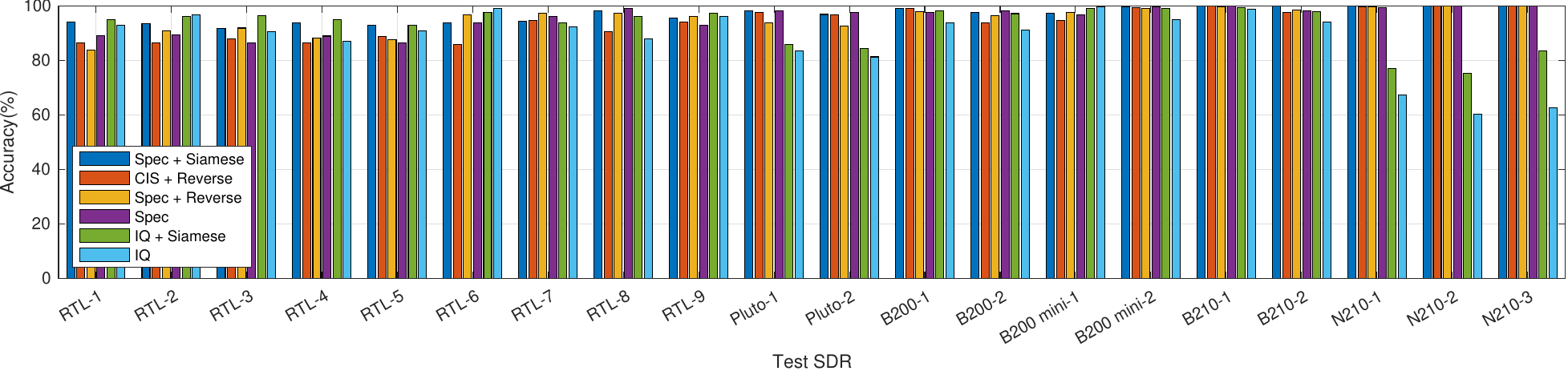}
		\caption{Evaluation of the proposed Siamese network-based RFFI for the receiver effect mitigation. \textit{Public Dataset 3}~\cite{shen2023towards} was used. The training dataset was collected by B200-1 and B210-1. The \textbf{test dataset A} was used.
        }
		\label{fig:rec_ago}
	\end{center}	
\end{figure*}
\mj{Section~\ref{sec:publicdata2_2} involves different receivers in the training and test stages.
Only the \textbf{test dataset A} with high SNR signals is used in this evaluation, to focus on the receiver effects. All the twenty SDR receivers were used.}

\mj{Fig.~\ref{fig:rec_ago} depicts the classification accuracy of various receivers with different methods. 
The proposed \textit{Spec + Siamese} achieves the best overall classification accuracy.
The performance of \textit{CIS + Reverse} and \textit{Spec + Reverse} is slightly worse than that of \textit{Spec + Siamese} in most cases. 
The occasional advantages are observed with the IQ-based methods, these occur in scenarios where all methods achieve relatively high classification accuracy, and the improvement brought by IQ-based methods is marginal. 
In contrast, in certain cases, the IQ-based method performs significantly worse than others, for instance, in scenarios involving receivers such as N210-1/2/3.
As previously mentioned, when IQ data is used as input, wireless channel effects are embedded in the received signal in a convolved manner, making it difficult to eliminate. As a result, it is more challenging for IQ-based approaches to maintain consistently high performance.}

\subsection{Evaluation Against Static Channel and Receiver Effects}
\label{sec:eva_rx_channel_static}
The \textbf{test dataset B} of the \textit{Public Dataset 3} presented in Section~\ref{sec:publicdata2_2} is used in this section, which involves different receivers in the training and test datasets as well as static multipath channel effects. Different from the \textbf{test dataset~A}, only three USRP N210 platforms were used in the \textbf{test dataset~B}.

\begin{figure}[!t]
\centering
\subfloat[]{\includegraphics[width=3.4in]{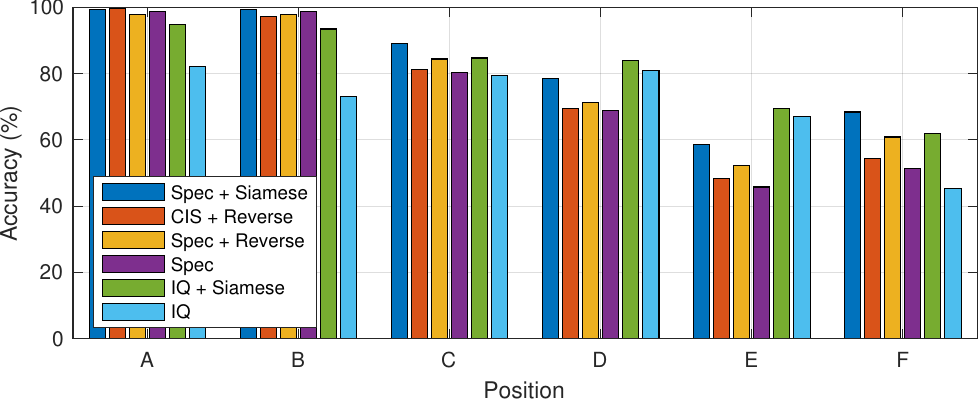}\label{fig:n210-1}}

\subfloat[]{\includegraphics[width=3.4in]{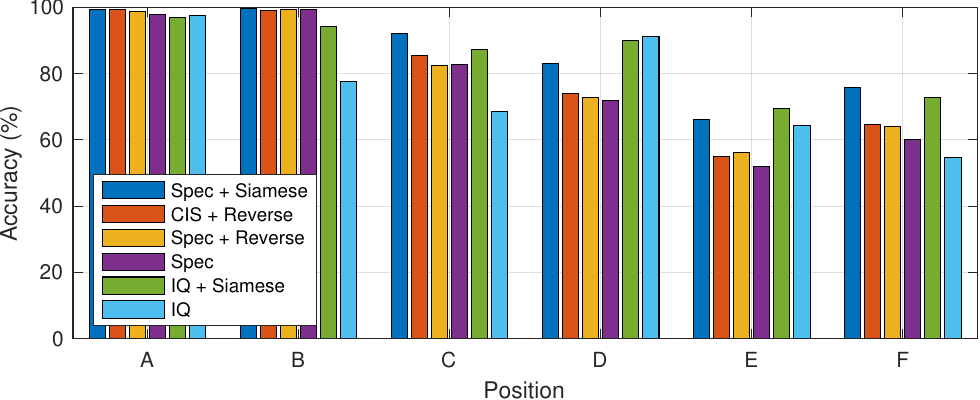}\label{fig:n210-2}}

\subfloat[]{\includegraphics[width=3.4in]{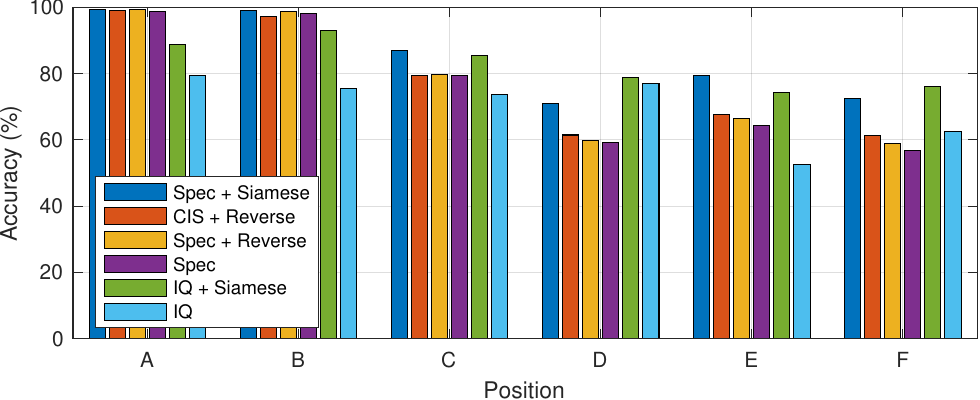}\label{fig:n210-3}}
\centering
\caption{Evaluation of the proposed Siamese network-based RFFI for the static channel and receiver effect mitigation. \textit{Public Dataset 3}~\cite{shen2023towards} was used. The training dataset was collected by B200-1 and B210-1. The \textbf{test dataset B} was used. The SNR of test datasets from A to F decreases from $50$ to $10$~dB.
(a) Test dataset from N210-1. (b) Test dataset from N210-2. (c) Test dataset from N210-3.}
\label{fig:exp_rx_channel_static}
\end{figure}

The classification results in various static channel environments are shown in Fig.~\ref{fig:exp_rx_channel_static}.
\textit{Spec + Siamese} demonstrates the best performance in most cases.
Nevertheless, it can be observed that in certain cases, such as position D, using IQ as the signal representation produced higher classification accuracies.
This might be due to that the channel effects in these positions have been augmented during the training stage. However, the IQ-based approaches do not have a consistently high classification accuracy. 

\subsection{Evaluation Against Dynamic Channel and Receiver Effects}\label{sec:eva_rx_channel_dynamic}
The dataset introduced in Section~\ref{sec:selfdata} is used in this section, which involves different receivers and dynamic channel effects.
The receivers used in the training dataset were B210-a and N210-c while the receivers used in the test dataset were RTL-a/b, B210-a/b and N210-a/b.

\begin{figure}[!t]
\centering
\subfloat[]{\includegraphics[width=3.4in]{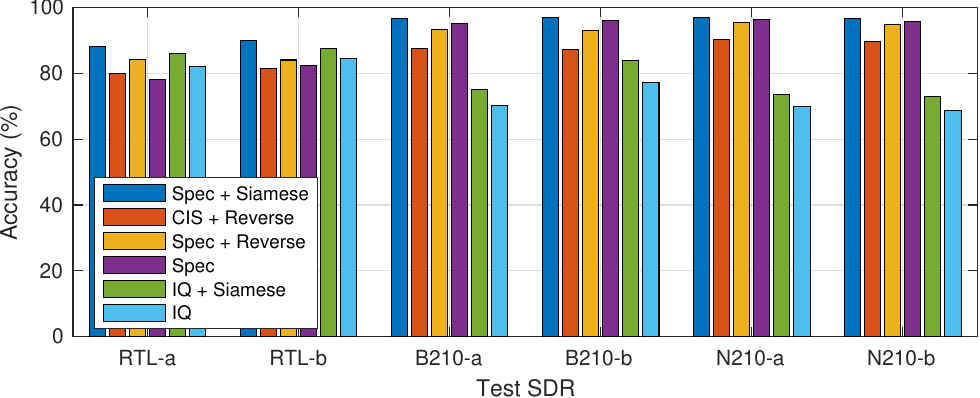}\label{fig:los-self}}

\subfloat[]{\includegraphics[width=3.4in]{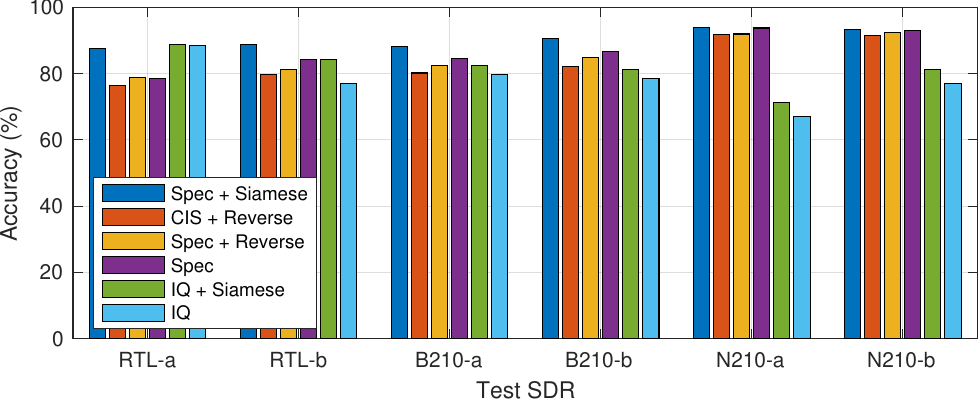}\label{fig:nlos-self}}
\centering
\caption{Evaluation of the proposed Siamese network-based RFFI for the dynamic channel and receiver effect mitigation. 
Training dataset was collected by B210-a and N210-c. (a) Test dataset collected under LOS scenario. (b) Test dataset collected under NLOS scenario.}
\label{fig:exp_rx_channel_dynamic}
\end{figure}

The classification accuracies are shown in Fig.~\ref{fig:exp_rx_channel_dynamic}. 
It can be found that the accuracy of \textit{Spec + Siamese} is over 85\% across all the receivers in both LOS and NLOS scenarios.
Its performance surpasses other approaches and demonstrates greater stability. In contrast, IQ samples demonstrated unstable performance again.
Fig.~\ref{fig:exp_rx_channel_dynamic} shows that utilizing a spectrogram as input yields higher accuracy compared to using IQ as input in most cases, which aligns with the observation in Fig.~\ref{fig:exp_rx_channel_static}.

\subsection{Complexity and Inference Time Comparison}
\mj{Table~\ref{tab:comparison} provides a comprehensive comparison of computational complexity and inference efficiency across different methods on four key metrics: inference time, number of parameters, storage requirements, and floating point operations (FLOPs).
Among all evaluated configurations, the methods with spectrogram or CIS show more efficient performance than IQ methods. 
The methods based on Spec or CIS maintain low inference times, approximately 1.13 to 1.28 milliseconds, and consistently require around 13 million parameters and less than 54 MB of storage.
In contrast, methods utilizing the IQ representation are substantially more resource-intensive.
\begin{table}[!t]
\caption{Computational Complexity and Inference Time}
\label{tab:comparison}
\begin{tabular}{|l|l|l|l|l|}
\hline
Method        & \begin{tabular}[c]{@{}l@{}}Inference \\ Time (ms)\end{tabular} & \begin{tabular}[c]{@{}l@{}}Parameter\\  (M)\end{tabular} & \begin{tabular}[c]{@{}l@{}}Storage \\ (MB)\end{tabular} & \begin{tabular}[c]{@{}l@{}}FLOPs \\ (G)\end{tabular} \\ \hline
Spec+ Siamese & 1.13                                                           & 13.94                                                    & 53.19                                                   & 0.58                                                 \\ \hline
CIS + Siamese & 1.25                                                           & 13.08                                                    & 49.94                                                   & 0.58                                                 \\ \hline
CIS + Reverse & 1.28                                                           & 13.09                                                    & 49.94                                                   & 0.58                                                 \\ \hline
Spec+ Reverse & 1.16                                                           & 13.94                                                    & 53.19                                                   & 0.58                                                 \\ \hline
Spec          & 1.13                                                           & 13.94                                                    & 53.19                                                   & 0.58                                                 \\ \hline
IQ + Siamese  & 2.82                                                           & 132.45                                                   & 505.26                                                  & 1.54                                                 \\ \hline
IQ            & 2.82                                                           & 132.45                                                   & 505.26                                                  & 1.54                                                 \\ \hline
\end{tabular}
\end{table}}
\subsection{Summary}
According to the above experimental evaluation, we can conclude that using the spectrogram demonstrated more consistent and higher classification accuracies, compared to the IQ samples. In addition, by adopting the Siamese network architecture, the classification performance can be further boosted. For example, as can be observed in Figs.~\ref{fig:exp_rx_channel_static} and ~\ref{fig:exp_rx_channel_dynamic}, \textit{Spec + Siamese} achieved about 10\% gain over \textit{Spec}.
Therefore, \textit{Spec + Siamese} is recommended and will be used for evaluating pretraining in Section~\ref{sec:eva_pre}.

\section{Experimental Evaluation of Pretraining}  \label{sec:eva_pre}
In this section, we used a pretraining dataset to train the feature extractor and then fine-tune the classification neural network, as shown in Fig.~\ref{fig:system}. We employed the \textit{Spec + Siamese} approach, as recommended in Section~\ref{sec:eva_siamese}. This three-stage approach is denoted as \textit{w/ pretrain}.
We also used training the \textit{Spec + Siamese} from scratch as the benchmark, denoted as \textit{w/o pretrain}, which is the two-stage approach evaluated in Section~\ref{sec:eva_siamese}.

Similar to Section~\ref{sec:eva_siamese}, we also did ablation studies for the effect of pretraining against channel effect (Section~\ref{sec:eva_pretrain_channel}) and receiver effect (Section~\ref{sec:eva_pretrain_rx}). We then investigated the scenarios when both the channel and receiver effects were present in
Sections~\ref{sec:pretrain_rx_channelstatic} and \ref{sec:pretrain_rx_channeldynamic}. For each evaluation, we carried out 4 rounds. The classification accuracy is presented in terms of the error bar.

\subsection{Setup for Pretraining} 
The datasets introduced in Section~\ref{sec:dataset} were also used in this section as the fine-tuning dataset. In addition, the feature extractor needs to be pretrained, and the dataset as well as the configuration will be introduced below.

\subsubsection{Dataset for Pretraining}\label{sec:predata}
The work in~\cite{shen2024federated} released a public LoRa dataset involving various types of DUTs and SDR receivers, detailed in Table~\ref{tab:dataset_training}. We used parts of the datasets in~\cite{shen2024federated} to build our pretraining dataset. 
In particular, we considered the signals from each receiver as a subdataset. We randomly selected a few DUTs from each subdataset to emulate real scenarios.
Specifically, the numbers of DUTs from the four receivers/subdatasets are 3/8/14/17, respectively. The number of packets obtained from each DUT ranges from $500$ to $800$. 
\begin{table}[!t]
\caption{Pretraining Dataset}
\label{tab:dataset_training}
\begin{tabular}{|lll||ll|}
\hline
\multicolumn{3}{|c||}{LoRa   Dataset in~\cite{shen2024federated}}                                                 & \multicolumn{2}{c|}{$\mathcal{D}_p$}                                          \bigstrut\\ \hline
\multicolumn{1}{|l|}{Rx}                                & \multicolumn{1}{l|}{DUT}              & DUT Indexes & \multicolumn{1}{l|}{Subset}                             & \# of DUT            \bigstrut\\ \hline
\multicolumn{1}{|l|}{Pluto}                             & \multicolumn{1}{l|}{Lopy4}            & 1 - 5          & \multicolumn{1}{l|}{$\mathcal{D}_p^1$}                  & 3                   \bigstrut\\ \hline
\multicolumn{1}{|l|}{\multirow{2}{*}{B200 mini}} & \multicolumn{1}{l|}{Lopy4}            & 6 - 10         & \multicolumn{1}{l|}{\multirow{2}{*}{$\mathcal{D}_p^2$}} & \multirow{2}{*}{8}  \\ \cline{2-3}
\multicolumn{1}{|l|}{}                                  & \multicolumn{1}{l|}{mbed 1272 shield} & 1 - 5          & \multicolumn{1}{l|}{}                                   &                     \bigstrut\\ \hline
\multicolumn{1}{|l|}{\multirow{3}{*}{B200}}      & \multicolumn{1}{l|}{Lopy4}            & 11 - 15        & \multicolumn{1}{l|}{\multirow{3}{*}{$\mathcal{D}_p^3$}} & \multirow{3}{*}{14} \\ \cline{2-3}
\multicolumn{1}{|l|}{}                                  & \multicolumn{1}{l|}{mbed 1272 shield} & 6 - 10         & \multicolumn{1}{l|}{}                                   &                     \\ \cline{2-3}
\multicolumn{1}{|l|}{}                                  & \multicolumn{1}{l|}{mbed 1261 shield} & 1 - 5          & \multicolumn{1}{l|}{}                                   &                     \bigstrut\\ \hline
\multicolumn{1}{|l|}{\multirow{4}{*}{B210}}      & \multicolumn{1}{l|}{Lopy4}            & 16 - 20        & \multicolumn{1}{l|}{\multirow{4}{*}{$\mathcal{D}_p^4$}} & \multirow{4}{*}{17} \\ \cline{2-3}
\multicolumn{1}{|l|}{}                                  & \multicolumn{1}{l|}{mbed 1272 shield} & 11 - 15        & \multicolumn{1}{l|}{}                                   &                     \\ \cline{2-3}
\multicolumn{1}{|l|}{}                                  & \multicolumn{1}{l|}{mbed 1261 shield} & 6 - 10         & \multicolumn{1}{l|}{}                                   &                     \\ \cline{2-3}
\multicolumn{1}{|l|}{}                                  & \multicolumn{1}{l|}{Fipy}             & 1 - 5          & \multicolumn{1}{l|}{}                                   &                     \bigstrut\\ \hline
\end{tabular}
\end{table}

\subsubsection{Architecture of Feature Extractor} \label{sec:trainconfig}
The feature extractor in Fig.~\ref{fig:resnet}, which has been described in Section~\ref{sec:dlarchitecture}, was trained in the pretraining stage.

\subsubsection{Pretraining Configuration} 
The parameters of the neural network were optimized by the Adam optimizer with an initial learning rate of $0.001$ and a batch size of $32$. 
The configurations for the learning rate scheduler and early stopping were identical to those of Section~\ref{sec:traincon}.

\subsection{Evaluation Against Channel Effect}\label{sec:eva_pretrain_channel}

The \textit{Public Dataset 1}~\cite{shen2022towards} in Section~\ref{sec:publicdata1} was used for stages 2 and 3. In particular, two test datasets, D1 and D4, were selected, representing LOS and NLOS conditions, respectively.

\begin{figure}[!t]
	\begin{center}
        \includegraphics[width = 3.4in]{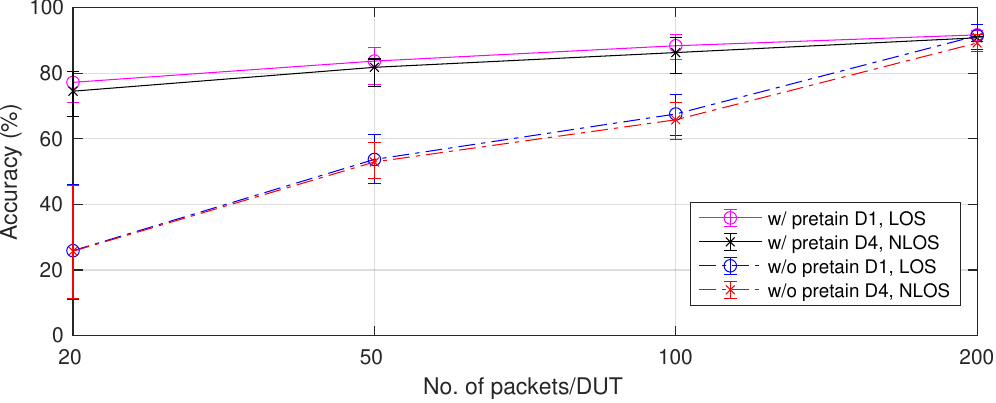}
		\caption{Evaluation of pretraining for mitigating channel effects. \textit{Public Dataset 1}~\cite{shen2022towards} was used. Test datasets D1 (LoS) and D4 (NLOS) were used.
		\label{fig:wowLOSNLOS}}
	\end{center}	
\end{figure}

The results are given in Fig.~\ref{fig:wowLOSNLOS}. 
It can be observed that the models trained with the pretraining (\textit{w/ pretrain}) show better performance than the models trained from scratch (\textit{w/o pretrain}), in both LOS and NLOS scenarios. 
The improvement in classification accuracy from the pretrained model is particularly evident when the number of data packets used for training is small. 
In the LOS scenario D1, the utilization of a pretrained feature extractor results in an improvement in classification accuracy from 25.9\% to 77.2\%,  when employing $20$ packets per DUT for fine-tuning. The improvement provided by the pretrained model gradually decreases when the number of fine-tuning data packets increases.
Both approaches achieved almost the same accuracies for LOS and NLOS scenarios when 200 fine-tuning data packets per DUT were used.

\subsection{Evaluation Against Receiver Effect}\label{sec:eva_pretrain_rx}
The \textit{Public Dataset 3}~\cite{shen2023towards} in Section~\ref{sec:publicdata2_2} was used. The training datasets from B200-1 and B210-1 were used for fine-tuning.
Datasets collected by RTL-1, Pluto-1, and N210-1 were chosen to form the test set for evaluation.

\begin{figure}[!t]
	\centering
	\subfloat[]{\includegraphics[width=1.15in]{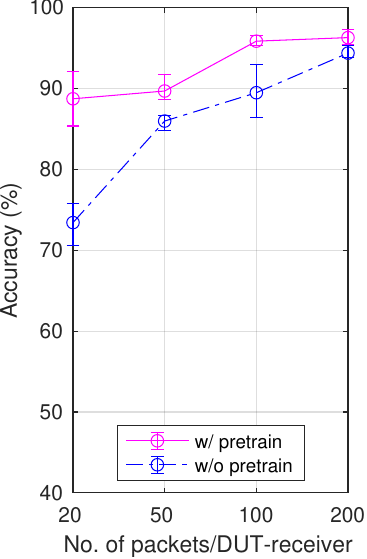}
		\label{fig:rtl}}
    \subfloat[]{\includegraphics[width=1.15in]{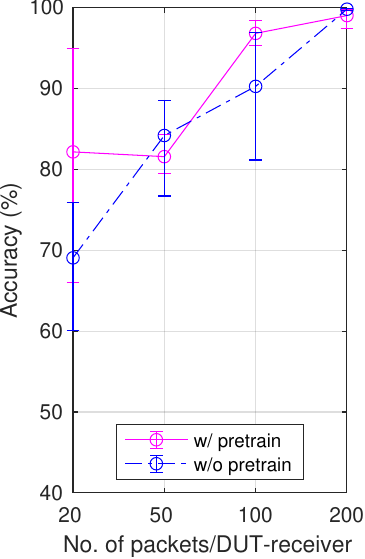}
 \label{fig:pluto}}	
     \subfloat[]{\includegraphics[width=1.15in]{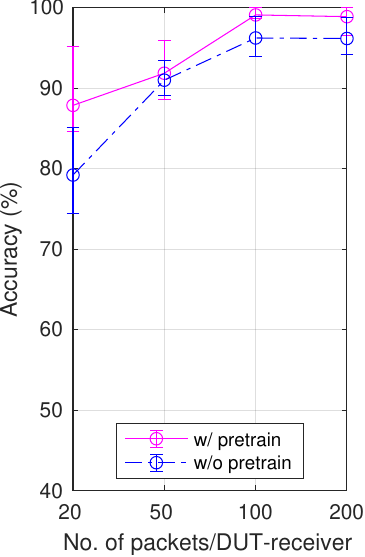}
      \label{fig:n210}}	
 \caption{Evaluation of pretrain on mitigating receiver effects. \textit{Public Dataset 3}~\cite{shen2023towards} was used. Training dataset from B200-1 and B210-1. Test datasets from static channels, half a meter between DUT and receiver. (a) Test dataset from RTL-1. (b) Test dataset from Pluto-1. (c) Test dataset from N210-1.}
	\label{fig:wowrtlpluton210}
\end{figure}

The impact of pretrained models on the receiver mitigation is presented in Fig.~\ref{fig:wowrtlpluton210}. When there were only 20 packets per DUT-receiver pair, the \textit{w/ pretrain} achieves an accuracy improvement of 10\% to 15\%, compared to \textit{w/o pretrain}. 
Taking RTL-1 as an example, as shown in Fig.~\ref{fig:wowrtlpluton210}(a), the average accuracy of \textit{w/o pretrain} is 74.5\%, while \textit{w/ pretrain} can achieve 88.7\%.
Therefore, pretraining is highly useful in getting a well-performing RFFI system when fine-tuning the network with few packets.

It can also be observed that as the number of training data packets increases, the impact of whether the pre-model is utilized gradually diminishes. When there were 200 packets per DUT-receiver pair, the results of \textit{w/ pretrain} and \textit{w/o pretrain} became nearly identical. 
This phenomenon exhibits consistency across multiple SDR platforms. 

\subsection{Evaluation Against Static Channel and Receiver Effects}\label{sec:pretrain_rx_channelstatic}
The \textit{Public Dataset 3}~\cite{shen2023towards} in Section~\ref{sec:publicdata2_2} was used. The training datasets from B200-1 and B210-1 were used for fine-tuning.
A subset from \textbf{test dataset B}, namely the data collected from Locations A, B and C using N210-1, was selected for evaluation.

\begin{figure}[!t]
	\centering
	\subfloat[]{\includegraphics[width=1.15in]{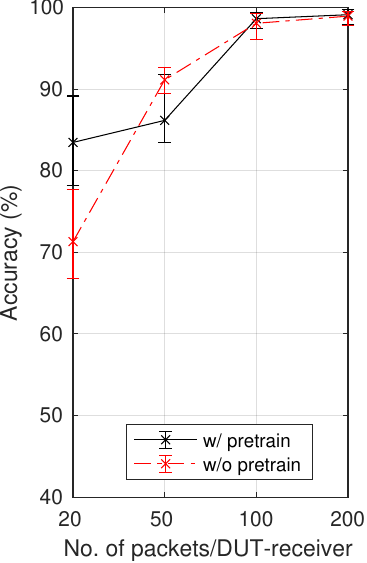}
		\label{fig:n2101a}}
            \subfloat[]{\includegraphics[width=1.15in]{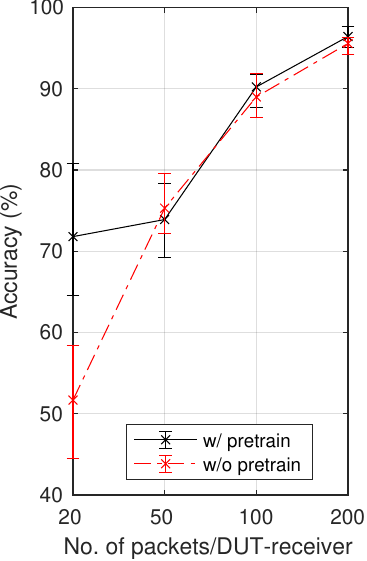}
 \label{fig:n210b}}	
    \subfloat[]{\includegraphics[width=1.15in]{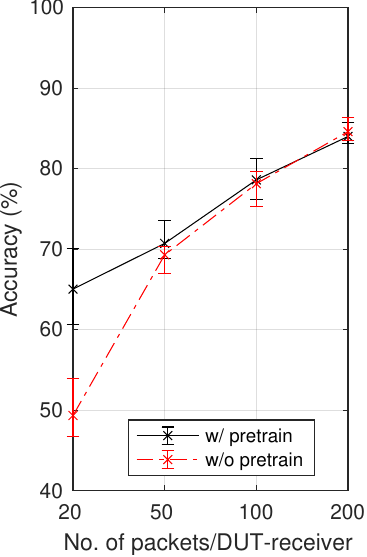}
 \label{fig:n210c}}	
 \caption{Evaluation of pretrain on mitigating static channel and receiver effects. \textit{Public Dataset 3}~\cite{shen2023towards} was used.  Training dataset from B200-1 and B210-1. Test datasets from static channels (NLOS) and collected by N210-1. (a) Test dataset from Location A. (b) Test dataset from Location B. (c) Test dataset from Location C.}
	\label{fig:wown2101abc}
\end{figure}

As can be observed in Fig.~\ref{fig:wown2101abc}, \textit{w/ pretrain} consistently achieves better performance across all channel environments when the number of fine-tuning samples was 20 per DUT-receiver, no matter Location A/B/C.
As shown in Fig.~\ref{fig:wown2101abc}(b), the accuracy in Location B improved by about 20\% with the pretrained model when only 20 packets were available for fine-tuning.

\subsection{Evaluation Against Dynamic Channel and Receiver Effects}\label{sec:pretrain_rx_channeldynamic}
The \textit{Self-Collected Dataset} in Section~\ref{sec:selfdata} was used. The training datasets from B210-a and N210-c were employed for fine-tuning.
The test dataset involving dynamic channel effects and receivers was used, with RTL-a, B210-a and N210-a chosen as examples.

\begin{figure}[!t]
	\centering
	\subfloat[]{\includegraphics[width=1.15in]{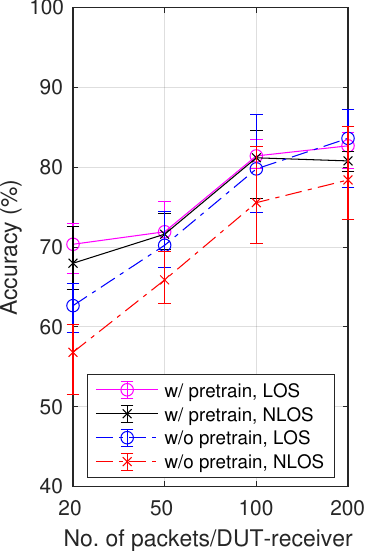}
		\label{fig:rtl_a}}
        \subfloat[]{\includegraphics[width=1.15in]{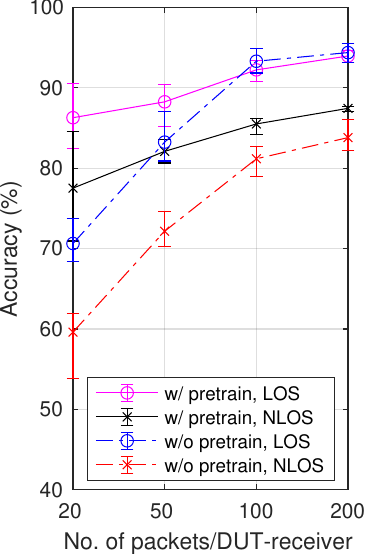}
            \label{fig:b210_a}}
    \subfloat[]{\includegraphics[width=1.15in]{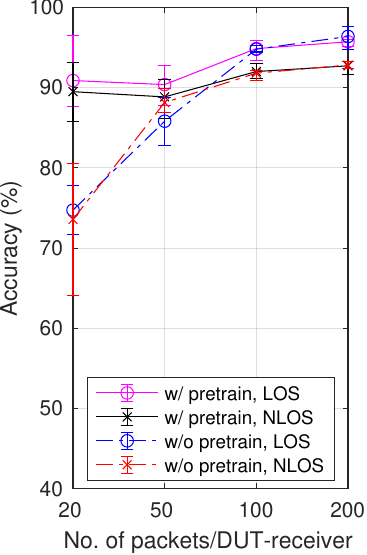}
            \label{fig:n210_a}}

	\caption{Evaluation of pretrain on mitigating dynamic channel and receiver effects. The \textit{Self-Collected Dataset} was used. Training dataset from B210-a and N210-c.  Test datasets from various receivers and dynamic channels.  (a) Test dataset from RTL-a. (b) Test dataset from B210-a. (c) Test dataset from N210-a. }
	\label{fig:self}
\end{figure}

Fig.~\ref{fig:self} shows experimental results in dynamic scenarios across multiple receivers.
The accuracy of all the receivers has been improved, both in LOS and NLOS, when the \textit{w/ pretrain} approach was used.
It can be found that across RTL-a, B210-a, and N210-a, \textit{w/ pretrain} boosted the accuracy by over 8\% on average when the number of training packets was $20$ from each DUT-receiver pair. 
More specifically, taking B210-a as an example, the classification accuracy increases to 86.3\%, surpassing the accuracy of 70.6\% by \textit{w/o pretrain}.

Fig.~\ref{fig:self}(c) illustrates that our proposed method enables the N210 to achieve above 90\% accuracy in dynamic NLOS scenarios when there is provided with only $20$ packets per DUT-receiver pair.
It also can be noted that, despite utilizing the same training packets, the accuracy of RTL is consistently lower than that of B210 and N210 in most cases. 
RTL is a low-cost receiver and its hardware characteristics are more variable, making it harder to eliminate its inherent characteristics.
Nonetheless, utilizing our proposed method, the classification accuracies of RTL-a/b are still improved. 
When provided with $100$ training packets per DUT-receiver pair, RTL-a achieves an accuracy of approximately 81.2\% in dynamic NLOS scenarios, demonstrating a 5.6\% increase compared to \textit{w/o pretrain}.

\subsection{Complexity  of the Three-stage Approach}
\begin{table}[!t]
\caption{Computational Complexity and Time Complexity of the Three-stage Approach}
\label{tab:totalcost}
\begin{tabular}{|l|l|l|l|l|}
\hline
\begin{tabular}[c]{@{}l@{}}Three-stage\\ Approach\end{tabular} & Time (ms)                                               & Epochs           & \begin{tabular}[c]{@{}l@{}}Parameter\\ (M)\end{tabular} & \begin{tabular}[c]{@{}l@{}}Storage\\ (M)\end{tabular} \\ \hline
\begin{tabular}[c]{@{}l@{}}Pretrain\end{tabular}    & $1.24\times10^7$                              & 175              & 13.93                                                   & 53.19                                                 \\ \hline
Finetune                                                       & $5\times10^4$$\sim$$7.43\times10^5$ & 68$\sim$129      & 13.94                                                   & 53.19                                                 \\ \hline
Inference                                                      & 1.13                                                    & \textbackslash{} & \textbackslash{}                                        & \textbackslash{}                                      \\ \hline
\end{tabular}
\end{table}
\mj{Table~\ref{tab:totalcost} presents the computational and time complexity of the proposed three-stage approach, comprising the pretraining phase for the feature extractor, the finetuning stage, and the inference stage. 
The pretrain stage is computationally intensive but only needs to be performed once.
The ``Finetune" stage shows a range for both time ($5\times10^4$ to $7.43\times10^5$ ms) and epochs (68 to 129), which corresponds to different training scenarios with varying packet-per-device configurations, specifically from 20 to 200 packets per device. 
The inference stage is highly efficient, requiring only 1.13 ms per sample.}

\subsection{Summary}
In general, \textit{w/ pretrain} has a better performance compared to \textit{w/o pretrain}, in terms of mean and variance of the classification accuracy. The benefits of \textit{w/ pretrain} are more significant when there are less packets for fune-tuning/training, because \textit{w/ pretrain} has a feature extractor pretrained.
In addition, the classification results exhibit significant fluctuations when the number of data packets is limited, characterized by the high error bar values. As the number of training data packets increases, the results become more stable, and the error bar values for each receiver decrease accordingly.
For example, Fig.~\ref{fig:self}(b) shows that the error bar value for B210-a under the LOS scenario is 8.2\% with $20$ training packets for each DUT-receiver pair, and it reduces to 1.5\% when increased to $200$ per DUT-receiver pair.

\section{Conclusion}\label{sec:conclusion}
\mj{In this paper, we proposed a three-stage RFFI approach involving a pretraining stage enhanced by contrastive learning and a Siamese network-based RFFI classification network to mitigate channel and receiver effects simultaneously. }
Firstly, spectrogram typically exhibits more stable and higher accuracy compared to IQ samples, which was adopted as the signal representation. 
Secondly, the Siamese network with contrastive loss was designed to mitigate performance degradation caused by varying channel effects and different receiver impairments. 
Finally, a pretraining approach using contrastive learning was proposed to address scenarios where only a limited number of labelled packets are available for training. 
We carried out extensive experimental evaluations using four different LoRa datasets, including three public LoRa datasets and one self-collected dataset.
With only 20 training packets per device, the accuracy can be significantly improved from 25.9\% to 77.2\% when utilizing a pretrained model.
The proposed RFFI system achieved a classification accuracy exceeding 90\% in distinguishing 10 LoRa devices across dynamic channels and various receivers. This paper considered LoRa-RFFI as a case study. However, our proposed three-stage approach is applicable to other wireless protocols. A new signal representation may need to be designed to expose signal features.

\bibliographystyle{IEEEtran}
\bibliography{IEEEabrv,mybibfile}
\end{document}